\documentclass[12pt,english,floatfix,nofootinbib,superscriptaddress,aps,prd,preprint]{revtex4}
\usepackage[utf8]{inputenc}
\usepackage{float}
\usepackage{array}
\usepackage{lipsum}
\usepackage{bbm}
\usepackage{dsfont}
\usepackage{graphicx}
\usepackage{amsmath}
\usepackage{tikz}
\usepackage{multirow}
\usepackage{braket}
\usepackage{bbm}
 \usetikzlibrary{quotes,angles}
 \usetikzlibrary{arrows} 
\usetikzlibrary{decorations.markings}
\usepackage{graphicx}
\usepackage[english]{babel}
\usepackage{color}
\usepackage{subfig}
\usepackage{caption}
\usepackage{tensor}
\usepackage{esint}
\usepackage[dvips]{epsfig}
\usepackage[dvips]{graphicx}
\usepackage{float}
\usepackage{units}
\usepackage{textcomp}
\usepackage{mathrsfs}
\usepackage{amsmath}
\usepackage[makeroom]{cancel}
\usepackage{amssymb}
\usepackage{amsbsy}
\usepackage{amsfonts}
\usepackage{amssymb,mathrsfs,xcolor}
\usepackage{esint}
\usepackage{array}
\usepackage{graphicx}

\usepackage{hyperref}
\hypersetup{
    colorlinks,
    citecolor=blue,
    filecolor=green,
    linkcolor=purple,
    urlcolor=red,
}

\usepackage{slashed}

\newcommand{\ie}{\begin{equation}}
\newcommand{\fe}{\end{equation}}
\newcommand{\se}{\begin{eqnarray}}
\newcommand{\ff}{\end{eqnarray}}

\usepackage{hyperref}
\hypersetup{colorlinks,breaklinks,
			citecolor=[rgb]{0,0.0,1.0},
            urlcolor=[rgb]{0.0,0.0,1.0},
            linkcolor=[rgb]{0,0.5,0.9}}

\begin{document}

\title{Particle production induced by a Lorentzian non--commutative spacetime}

\author{A. A. Ara\'{u}jo Filho}
\email{dilto@fisica.ufc.br}
\affiliation{Departamento de Física, Universidade Federal da Paraíba, Caixa Postal 5008, 58051--970, João Pessoa, Paraíba,  Brazil.}
\affiliation{Departamento de Física, Universidade Federal de Campina Grande Caixa Postal 10071, 58429-900 Campina Grande, Paraíba, Brazil.}

%%%%%%%%%%%%%%%%%%%%%%%%%%%%%%%%%%%%%%%%%%%%%%%%%%%%%%%%%%%%%%%%%%%%%%%%%%%%%%%%%%%%%%%%%%%%%%%%%%%%%%%%%%%%%%%%%%%%%%%%%%%%%%%%%%%%%%%%%%%%%%%%%%%%%%%%%%%%%%%%%%%%%%%%%%%%%%%%%%%%%%%%%%%%%%%%%%%%%%%%%%%%%%%%%%%%%%%%%%%%%%%%%%%%%%%%%%%%%%%%%%%%%%%%%%%%%%%%%%%%%%%%%%%%%%%%%%%%%%%%%%%%%%%%%%%%%%%%%%%%%%%%%%%%%%%%%%%%%%%%%%%%%%%%%%%%%%%%%%%%%%%%%%%%%%%%%%%%%%%%%%%%%%%%%%%%%%%%%%%%%%%%%%%%%%%%%%%

\date{\today}

\begin{abstract}

In this paper, we examine particle production, evaporation, and greybody factors for a Lorentzian non--commutative black hole. We begin by analyzing particle creation for bosons, considering scalar perturbations to compute the Bogoliubov coefficients, which enable the determination of the Hawking temperature $T_{\Theta}$. Subsequently, we describe Hawking radiation as a tunneling process using the Painlevé--Gullstrand metric representation, allowing the evaluation of divergent integrals via the residue method. This approach yields the particle creation density for bosonic modes. Next, we extend the analysis to fermions, obtaining the corresponding particle creation density. The black hole evaporation is then examined through the \textit{Stefan--Boltzmann} law, leading to an estimate of the black hole's lifetime. In this context, we identify the presence of a remnant mass when the black hole reaches the final stage of its evaporation. Furthermore, we compute greybody factors for bosons, taking into account scalar, vector, and tensorial perturbations. Finally, we determine the greybody factors for fermions as well. Overall, compared to the Schwarzschild case ($\Theta = 0$), the presence of the non--commutative parameter $\Theta$ lowers the Hawking temperature and reduces the particle creation density for both bosons and fermions, causing the evaporation process to proceed more slowly. Additionally, $\Theta$ decreases the magnitude of the greybody factors for bosons and fermions across all perturbations considered in this analysis.

\end{abstract}

%%\keywords{Thermodynamics properties; Higher-Derivative, Poldolsky eletrodynamics.}

\maketitle

\tableofcontents
%%%%%%%%%%%%%%%%%%%%%%%%%%%%%%%%%%%%%%%%%%%%%%%%%%%%%%%%%%%%%%%%%%%%%%%%%%%%%%%%%%%%%%%%%%%%%%%%%%%%%%%%%%%%%%%%%%%%%%%%%%%%%%%%%%%%%%%%%%%%%%%%%%%%%%%%%%%%%%%%%%%%%%%%%%%%%%%%%%%%%%%%%%%%%%%%%%%%%%%%%%%%%%%%%%%%%%%%%%%%%%%%%%%%%%%%%%%%%%%%%%%%%%%%%%%%%%%%%%%%%%%%%%%%%%%%%%%%%%%%%%%%%%%%%%%%%%%%%%%%%%%%%%%%%%%%%%%%%%%%%%%%%%%%%%%%%%%%%%%%%%%%%%%%%%%%%%%%%%%%%%%%%%%%%%%%%%%%%%%%%%%%%%%%%%%%%%%%%%%%%%%%%%%%%%%%%%%%%%%%%%%%%%%%%%%%%%%%%%%%%%%%%%%%%%%%%%%%%%%%%%%%%%%%%%%%%%%%%%%%%%%%%%

\pagebreak

\section{Introduction}

General relativity does not impose an absolute lower bound on measurable distances within spacetime. Nevertheless, the Planck length is often regarded as a fundamental threshold, hinting at a possible limitation to classical geometric descriptions. To account for such constraints, non--commutative spacetime models have been formulated, offering a framework that integrates quantum gravitational effects. These models, which are closely related to string theory and other approaches to quantum gravity, have become increasingly relevant in supersymmetric field theories, particularly when analyzed through the superfield formalism \cite{witten,szabo2003quantum}.

An effective strategy for incorporating non--commutative structures into gravitational theories involves the Seiberg--Witten map, which enables the gauging of symmetry groups in deformed field theories \cite{chamseddine2001deforming}. This formalism has been widely applied in black hole physics, providing a means to explore thermodynamic properties, evaporation dynamics, and thermal behavior. Research in this direction has investigated emission spectra, equilibrium configurations, and corrections to thermodynamic quantities, taking into account quantum effects in curved spacetime \cite{nozari2006reissner,lopez2006towards,touati2023thermodynamic,AraujoFilho:2024rss,sharif2011thermodynamics,nozari2007thermodynamics,banerjee2008noncommutative,t29,myung2007thermodynamics}.

Significant progress has been made in incorporating non--commutative effects into gravitational models by modifying the matter content in Einstein’s field equations while keeping the Einstein tensor unchanged \cite{nicolini2006noncommutative}. Traditional point--mass representations have been replaced with smoothly distributed density functions, such as Gaussian \cite{ghosh2018noncommutative} and Lorentzian profiles \cite{nicolini2009noncommutative}, offering a more refined approach to describing spacetime geometry. These adjustments have driven extensive investigations into black hole thermodynamics, particularly in the context of quantum tunneling and thermal radiation emission \cite{nozari2008hawking,banerjee2008noncommutative,sharif2011thermodynamics}.

Beyond thermodynamics, the implications of non--commutative modifications have been explored in various gravitational scenarios. Studies have examined their influence on topological features in Gauss--Bonnet gravity \cite{lekbich2024optical}, as well as their role in shaping geodesic motion \cite{touati2022geodesic}. The effects of non--commutativity have also been investigated in the context of black hole shadows \cite{sharif2016shadow,wei2015shadow,ovgun2020shadow}, matter accretion, and gravitational lensing phenomena \cite{ding2011strong,ding2011probing,saleem2023observable}. More recently, perturbative approaches have been employed to assess how non--commutativity introduces corrections in gravitational frameworks, broadening its theoretical relevance \cite{newcommutativity}.

\pagebreak

Quantum mechanics and gravity converge in Hawking’s groundbreaking discovery, which has played a crucial role in shaping the field of quantum gravity \cite{o111,o11,o1}. His work revealed that black holes are not entirely black but instead emit thermal radiation, leading to a gradual decrease in their mass. This process, now known as Hawking radiation, emerges from quantum field effects in curved spacetime near the event horizon \cite{parikh2004energy,eeeOvgun:2019jdo,eeeKuang:2018goo,eeeOvgun:2015box,eeeOvgun:2019ygw,gibbons1977cosmological,almheiri2021entropy,eeeKuang:2017sqa}. The implications of this phenomenon have profoundly influenced black hole thermodynamics and quantum effects in gravitational settings \cite{o7,aa2024implications,araujo2024dark,o9,sedaghatnia2023thermodynamical,o3,araujo2024charged,o6,araujo2023analysis,o8,o4}. An alternative perspective on Hawking radiation was later introduced by Kraus and Wilczek \cite{o10} and further refined by Parikh and Wilczek \cite{o12,o11,013}, framing the process as quantum tunneling across the event horizon. This semi--classical interpretation has since been widely applied to various black hole models \cite{mitra2007hawking,silva2013quantum,medved2002radiation,anacleto2015quantum,zhang2005new,mirekhtiary2024tunneling,del2024tunneling,senjaya2024bocharova,johnson2020hawking,vanzo2011tunnelling,5vanzzo,touati2024quantum,calmet2023quantum}.

This study explores particle production, evaporation, and greybody factors in the context of a Lorentzian non--commutative black hole. The analysis begins with bosonic particle creation, where scalar perturbations are introduced to compute the Bogoliubov coefficients, leading to the determination of the modified Hawking temperature $T_{\Theta}$. Hawking radiation is then examined through a tunneling framework, employing the Painlevé--Gullstrand metric representation. This formulation allow us to evaluate the divergent integrals via the residue method, providing the particle creation density for bosonic modes. The investigation is then extended to fermionic fields, yielding the corresponding particle production density as well. The black hole’s evaporation process is analyzed using the \textit{Stefan--Boltzmann} law, which allows for an estimation of its lifetime. In this framework, a remnant mass is identified as the black hole approaches the final stage of evaporation. Greybody factors for bosonic fields are subsequently computed, incorporating scalar, vector, and tensorial perturbations, followed by a similar evaluation for fermionic fields.

%%%%%%%%%%%%%%%%%%%%%%%%%%%%%%%%%%%%%%%%%%%%%%%%%%%%%%%%%%%%%%%%%%%%%%%%%%%%%%%%%%%%%%%%%%%%%%%%%%%%%%%%%%%%%%%%%%%%%%%%%%%%%%%%%%%%%%%%%%%%%%%%%%%%%%%%%%%%%%%%%%%%%%%%%%%%%%%%%%%%%%%%%%%%%%%%%%%%%%%%%%%%%%%%%%%%%%%%%%%%%%%%%%%%%%%%%%%%%%%%%%%%%%%%%%%%%%%%%%%%%%%%%%%%%%%%%%%%%%%%%%%%%%%%%%%%%%%%%%%%%%%%%%%%%%%%%%%%%%%%%%%%%%%%%%%%%%%%%%%%%%%%%%%%%%%%%%%%%%%%%%%%%%%%%%%%%%%%%%%%%%%%%%%%%%%%%%%%%%%%%%%%%%%%%%%%%%%%%%%%%%%%%%%%%%
\section{The black hole solution}

Integrating non--commutativity principles with general relativity allows for modifications to the structure of spacetime \cite{Anacleto:2019tdj,anacleto2023absorption,k10,k101,k102,k103,k6,k7,k8,k9,campos2022quasinormal}. Several formulations of non--commutative field theory have been developed based on the Moyal product \cite{k11}. This section begins by analyzing the essential properties of the black hole solution under consideration, starting with the given mass distribution function \cite{nozari2008hawking,nicolini2006noncommutative,campos2022quasinormal}, $  
\rho_{\Theta}(r) = \frac{M \sqrt{\Theta}}{\pi^{3/2} (r^{2} + \pi \Theta)^{2}}$, where $M$ represents the total mass, and $\Theta$ is the non--commutative parameter with dimensions of $[\mathrm{L}^{2}]$, defined through the commutation relation: 
$[x^{\mu},x^{\nu}] = i \Theta^{\mu \nu}$.  To describe how mass is distributed within this framework, the function $\mathcal{M}_{\Theta}$ is introduced and computed as  
$\mathcal{M}_{\Theta} = \int^{r}_{0} 4\pi r^{2}\rho_{\Theta}(r) \mathrm{d}r = M - \frac{4M \sqrt{\Theta}}{\sqrt{\pi}r}$.
With this formulation, a Schwarzschild--like black hole solution emerges in the non--commutative scenario \cite{Anacleto:2019tdj,campos2022quasinormal}
\ie
\mathrm{d}s^{2} = -A_{\Theta} (r) \mathrm{d}\tau^{2} + \frac{1}{B_{\Theta}(r)} \mathrm{d}r^{2} + r^{2}\mathrm{d}\theta^{2} + r^{2}\sin^{2}\theta \mathrm{d}\varphi^{2},
\fe
where
\ie
\label{huashuas}
A_{\Theta}(r) = B_{\Theta}(r)^{-1}= 1- \frac{2M}{r} + \frac{8M\sqrt{\Theta}}{\sqrt{\pi}r^{2}}.
\fe
By considering $1/g_{rr}=0$, we obtain the following solutions
\ie
r_{+} = M + \frac{\sqrt{\pi  M^2-8 \sqrt{\pi } \sqrt{\Theta } M}}{\sqrt{\pi }},
\fe
and
\ie
r_{-} = M-\frac{\sqrt{\pi  M^2-8 \sqrt{\pi } \sqrt{\Theta } M}}{\sqrt{\pi }},
\fe
in which $r_{+}$ and $r_{-}$ represent the event the Cauchy horizons, respectively.

%%%%%%%%%%%%%%%%%%%%%%%%%%%%%%%%%%%%%%%%%%%%%%%%%%%%%%%%%%%%%%%%%%%%%%%%%%%%%%%%%%%%%%%%%%%%%%%%%%%%%%%%%%%%%%%%%%%%%%%%%%%%%%%%%%%%%%%%%%%%%%%%%%%%%%%%%%%%%%%%%%%%%%%%%%%%%%%%%%%%%%%%%%%%%%%%%%%%%%%%%%%%%%%%%%%%%%%%%%%%%%%%%%%%%%%%%%%%%%%%%%%%%%%%%%%%%%%%%%%%%%%%%%%%%%%%%%%%%%%%%%%%%%%%%%%%%%%%%%%%%%%%%%%%%%%%%%%%%%%%%%%%%%%%%%%%%%%%%%%%%%%%%%%%%%%%%%%%%%%%%%%%%%%%%%%%%%%%%%%%%%%%%%%%%%%%%%%%%%%%%%%%%%%%%%%%%%%%%%%%

\section{Particle creation}

%%%%%%%%%%%%%%%%%%%%%%%%%%%%%%%%%%%%%%%%%%%%%%%%%%%%%%%%%%%%%%%%%%%%%%%%%%%%%%%%%%%%%%%%%%%%%%%%%%%%%%%%%%%%%%%%%%%%%%%%%%%%%%%%%%%%%%%%%%%%%%%%%%%%%%%%%%%%%%%%%%%%%%%%%%%%%%%%%%%%%%%%%%%%%%%%%%%%%%%%%%%%%%%%%%%%%%%%%%%%%%%%%%%%%%%%%%%%%%%%%%%%%%%%%%%%
\subsection{Bosonic case}

This study investigates how the presence of the non--commutative parameter $\Theta$ affects the emission of Hawking radiation. The analysis is motivated by Hawking’s seminal work \cite{hawking1975particle}, where he examined the quantum behavior of a scalar field in a curved spacetime background. His approach involved solving the wave equation for the scalar field, expressing the wave function as
\ie
\frac{1}{\sqrt{-g_{\Theta}}}\partial_{\mu}(g_{\Theta}^{\mu\nu}\sqrt{-g_{\Theta}}\partial_{\nu}\Phi) = 0.
\fe
Here, the inverse metric tensor is denoted by $g_{\Theta}^{\mu\nu}$, while $g_{\Theta}$ represents the determinant of the metric; and the scalar field is labeled as $\Phi$. In this manner, the corresponding field operator can be expressed as
\ie
\Phi = \sum_{i} \left (f^{\Theta}_i  a^{\Theta}_{i} + \bar{f}^{\Theta}_{i}  a^{{\Theta}\dagger}_{i} \right) = \sum_{i} \left( p^{\Theta}_{i}  b^{\Theta}_{i} + \bar{p}^{\Theta}_{i}  b^{{\Theta}\dagger}_{i} + q^{\Theta}_{i}  c^{\Theta}_{i} + \bar{q}^{\Theta}_{i}  c^{{\Theta}\dagger}_{i} \right ) .
\fe

In this framework, the functions $f^{\Theta}_{i}$ and $\bar{f}^{\Theta}_{i}$ (where $\bar{f}^{\Theta}_{i}$ denotes the complex conjugate) correspond to purely ingoing wave components. The solutions $p^{\Theta}_{i}$ and $\bar{p}^{\Theta}_{i}$ are associated exclusively with outgoing waves, whereas $q^{\Theta}_{i}$ and $\bar{q}^{\Theta}_{i}$ describe modes that lack outgoing contributions. The operators $a^{\Theta}_{i}$, $b^{\Theta}_{i}$, and $c^{\Theta}_{i}$ function as annihilation operators, while their conjugate counterparts $a^{\Theta \dagger}_{i}$, $b^{\Theta \dagger}_{i}$, and $c^{\Theta \dagger}_{i}$ serve as creation operators.  The focus here is to determine how non--commutativity influences these wave solutions. In other words, the objective is to examine the extent to which the parameter $\Theta$ modifies Hawking’s original formulation, altering the behavior of $f^{\Theta}_{i}$, $\bar{f}^{\Theta}_{i}$, $p^{\Theta}_{i}$, $\bar{p}^{\Theta}_{i}$, $q^{\Theta}_{i}$, and $\bar{q}^{\Theta}_{i}$.

Given the spherical symmetry of the metric under investigation, the wave solutions for both ingoing and outgoing modes can be expressed using spherical harmonics. In the region outside the black hole horizon, these solutions take the following form \cite{calmet2023quantum,heidari2024quantum,touati2024quantum}:
\begin{eqnarray}
f^{\Theta}_{\omega^\prime l m} &=& \frac{1}{\sqrt{2 \pi \omega^\prime} r }  \mathcal{F}_{\omega^{\prime}}^{\Theta}(r) e^{i \omega^\prime v^{\Theta}} Y_{lm}(\theta,\phi)\ , \\ 
p^{\Theta}_{\omega l m} &=& \frac{1}{\sqrt{2 \pi \omega} r }  \mathcal{P}^{\Theta}_\omega(r) e^{i \omega u^{\Theta}} Y_{lm}(\theta,\phi). 
\end{eqnarray}
In this formulation, the radial functions are represented by $\mathcal{F}_{\omega^{\prime}}^{\Theta}(r)$ and $\mathcal{P}^{\Theta}_\omega(r)$, while the angular dependence is captured through the spherical harmonics $Y_{lm}(\theta,\phi)$. The coordinates $v^{\Theta}$ and $u^{\Theta}$, corresponding to advanced and retarded coordinates, are defined as $
v^{\Theta} = t + r^{*}, \quad u^{\Theta} = t - r^{*}$,  
where $r^{*}$ denotes the tortoise coordinate.

With these definitions in place, the objective is to identify the corrections introduced by non--commutativity in the coordinate functions. A useful way to approach this problem is by analyzing the motion of a particle traveling along a geodesic within the modified spacetime, where its trajectory is parameterized by an affine parameter $s$. Within this setup, the particle’s momentum takes the form
\ie
p^{\Theta}_{\mu} = g^{\Theta}_{\mu\nu}\frac{\mathrm{d}x}{\mathrm{d}s}^\nu.
\fe
The momentum remains a conserved quantity along the geodesic trajectory. Additionally, the corresponding expression can be written as
\ie
\mathcal{L} = g^{\Theta}_{\mu\nu} \frac{\mathrm{d}x^\mu}{\mathrm{d}s} \frac{\mathrm{d}x^\nu}{\mathrm{d}s},
\fe
where this quantity also remains conserved along geodesic paths. For massive particles, we set $\mathcal{L} = -1$ and choose $s = \tau$, where $\tau$ represents the proper time. In contrast, for massless particles—the primary focus of this analysis --- we assign $\mathcal{L} = 0$, with $s$ acting as a general affine parameter.  

Notice that if we regard a stationary, spherically symmetric spacetime and restricting the analysis to radial geodesics, where $p^{\Theta}_\varphi = L = 0$, within the equatorial plane $(\theta = \pi/2)$ approach, the relevant equations governing the motion can be determined
\ie
E =  A_{\Theta}(r) \dot{t}.
\fe
In this formulation, the particle’s energy is defined as $E = -p^{\Theta}_{t}$, while differentiation with respect to the affine parameter $s$ is denoted by a dot, representing $\mathrm{d}/\mathrm{d}s$. Following this, an additional relation can be established, leading to
\ie
\label{1d2r3d4s}
    \left( \frac{\mathrm{d}r}{\mathrm{d}s} \right)^2 = \frac{E^2}{A_{\Theta}(r)B_{\Theta}(r)^{-1}},
\fe
and through a series of algebraic transformations, the following expression is obtained
\ie
\label{1c2o3n4ccc}
    \frac{\mathrm{d}}{\mathrm{d}s}\left(t\mp r^{*}\right) = 0,
\fe
where the tortoise coordinate $r^{*}$ is defined below
\ie
\begin{split}
\label{1t2o3r4t5o6ise}
\mathrm{d}r^{*} & = \frac{\mathrm{d}r}{\sqrt{A_{\Theta}(r)B_{\Theta}(r)}},
\end{split}
\fe
or, explicitly, it is written as
\ie
\label{rstar}
\begin{split}
& r^{*}= r \, + \, M \ln \left(8 \sqrt{\Theta } M+\sqrt{\pi } r (r-2 M)\right)\\
& +  \frac{2 \sqrt{M} \left(\sqrt{\pi } M-4 \sqrt{\Theta }\right) \tanh ^{-1}\left(\frac{\sqrt[4]{\pi } (M-r)}{\sqrt{M} \sqrt{\sqrt{\pi } M-8 \sqrt{\Theta }}}\right)}{\sqrt[4]{\pi } \sqrt{\sqrt{\pi } M-8 \sqrt{\Theta }}}.
\end{split}
\fe

In addition, it is important to observe that Eq. (\ref{1c2o3n4ccc}) gives rise to two conserved quantities, namely $v^{\Theta}$ and $u^{\Theta}$. By manipulating the expression for the retarded coordinate, an alternative formulation is obtained:
\ie
\label{ghjghk}
\frac{\mathrm{d}u^{\Theta}}{\mathrm{d}s}=\frac{2E}{A_{\Theta}(r)}.
\fe

For an ingoing geodesic parameterized by $s$, the advanced coordinate $u^{\Theta}$ is treated as a function of $s$, denoted as $u^{\Theta}(s)$. To determine its exact form, two fundamental steps are required: rewriting the radial coordinate $r$ as a function of $s$, and, subsequently, performing the integration outlined in Eq. \eqref{ghjghk}. The final expression for $u^{\Theta}(s)$ directly influences the Bogoliubov coefficients, which are essential in characterizing the quantum radiation emitted by the black hole.  

To move forward with this derivation, the functions $A_{\Theta}(r)$ and $B_{\Theta}(r)$ are used to perform the integration of the square root term appearing in Eq. \eqref{1d2r3d4s}. This integration is carried out over the range $\tilde{r} \in [r_{+}, r]$, while the affine parameter spans $\tilde{s} \in [0, s]$. Implementing this procedure, the resulting expression is obtained as follows:
\ie
r(\Theta,s) =  M + \frac{\sqrt{\pi  M^2-8 \sqrt{\pi } \sqrt{\Theta } M}}{\sqrt{\pi }} - E s.
\fe
Notice that, to arrive at this result, the negative sign in the square root was selected when solving Eq. \eqref{1d2r3d4s}, ensuring consistency with the ingoing geodesic trajectory.

Proceeding further, the function $r(s, \Theta)$ is employed to perform the integration. By focusing on the region near the event horizon \cite{parker2009quantum}, the following result is written 
\ie
\begin{split}
& u_{\text{vicinity}}^{\Theta}(s) \approx -  \frac{1}{{\pi ^{3/4} M-8 \sqrt[4]{\pi } \sqrt{\Theta }}} \\
& \times    \left[2 \ln \left(\frac{s}{C}\right) \left(\pi ^{3/4} M^2-8 \sqrt[4]{\pi } \sqrt{\Theta } M+\sqrt{\pi } M \sqrt{M \left(\sqrt{\pi } M-8 \sqrt{\Theta }\right)} \right.\right. \\
& \left.\left. -4 \sqrt{\Theta } \sqrt{M \left(\sqrt{\pi } M-8 \sqrt{\Theta }\right)}\right)\right],
\end{split}
\fe
where $C$ represents an integration constant. Additionally, in the asymptotic region far from the event horizon \cite{parker2009quantum}, the expression takes the form
\ie
\begin{split}
u^{\Theta}_{far}(s) \approx & \; 2 E \lambda -\frac{2 \sqrt{M} \left(-4 \sqrt{\Theta }-\sqrt[4]{\pi } \sqrt{M \left(\sqrt{\pi } M-8 \sqrt{\Theta }\right)}+\sqrt{\pi } M\right) \tanh ^{-1}\left(\frac{\sqrt[4]{\pi } E \lambda }{2 \sqrt{M} \sqrt{\sqrt{\pi } M-8 \sqrt{\Theta }}}\right)}{\sqrt[4]{\pi } \sqrt{\sqrt{\pi } M-8 \sqrt{\Theta }}}\\
& \frac{1}{{\pi ^{3/4} M-8 \sqrt[4]{\pi } \sqrt{\Theta }}} \times \left\{   \left(\pi ^{3/4} M^2-M \left(8 \sqrt[4]{\pi } \sqrt{\Theta }+\sqrt{\pi } \sqrt{M \left(\sqrt{\pi } M-8 \sqrt{\Theta }\right)}\right) \right.\right. 
\\
& \left.\left. +4 \sqrt{\Theta } \sqrt{M \left(\sqrt{\pi } M-8 \sqrt{\Theta }\right)}\right) \ln \left(-\sqrt{\pi } E^2 \lambda ^2+4 \sqrt{\pi } M^2-32 \sqrt{\Theta } M\right) \right\}.
\end{split}
\fe
However, when analyzing particle creation, only $ u_{\text{vicinity}}^{\Theta}(s)$ will be considered, following the standard approach adopted in the literature \cite{parker2009quantum,calmet2023quantum,touati2024quantum,araujo2025remarks,aa2024particle,aa2025does}. Additionally, the relationship between ingoing and outgoing null coordinates can be understood through the principles of geometric optics. This connection is expressed in terms of the parameter $s$, which satisfies the equation  $
s = \frac{v^{\Theta}_{0} - v^{\Theta}}{D}$,
where $v^{\Theta}_0$ corresponds to the advanced coordinate at the horizon reflection point ($s = 0$), and $D$ represents a constant factor \cite{calmet2023quantum}.

Having laid the groundwork, the next step is to determine the outgoing solutions of the modified Klein--Gordon equation while accounting for the effects introduced by the non--commutative parameter $\Theta$. The obtained expressions take the following form:
\ie
p^{\Theta}_{\omega} =\int_0^\infty \left ( \alpha^{\Theta}_{\omega\omega^\prime} f^{\Theta}_{\omega^\prime} + \beta^{\Theta}_{\omega\omega^\prime} \bar{ f}^{\Theta}_{\omega^\prime}  \right)\mathrm{d} \omega^\prime,
\fe
where $\alpha^{\Theta}_{\omega\omega^\prime}$ and $\beta^{\Theta}_{\omega\omega^\prime}$ represent the Bogoliubov coefficients, which characterize particle creation and mode mixing in the presence of non--commutativity \cite{parker2009quantum,hollands2015quantum,wald1994quantum,fulling1989aspects}
\begin{equation}
\begin{split}
\alpha^{\Theta}_{\omega\omega^\prime} = & -i K e^{i\omega^\prime v^{\Theta}_{0}}e^{\pi \left\{  \frac{1}{{\pi ^{3/4} M-8 \sqrt[4]{\pi } \sqrt{\Theta }}}     \left[ \left(\pi ^{3/4} M^2-8 \sqrt[4]{\pi } \sqrt{\Theta } M+\sqrt{\pi } M \sqrt{M \left(\sqrt{\pi } M-8 \sqrt{\Theta }\right)}  -4 \sqrt{\Theta } \sqrt{M \left(\sqrt{\pi } M-8 \sqrt{\Theta }\right)}\right)\right] \right\} \omega} \\
& \times \int_{-\infty}^{0} \,\mathrm{d}x\,\Big(\frac{\omega^\prime}{\omega}\Big)^{1/2}e^{\omega^\prime x} \\
& \times e^{i\omega\left\{ \frac{1}{{\pi ^{3/4} M-8 \sqrt[4]{\pi } \sqrt{\Theta }}}     \left[ 2 \left(\pi ^{3/4} M^2-8 \sqrt[4]{\pi } \sqrt{\Theta } M+\sqrt{\pi } M \sqrt{M \left(\sqrt{\pi } M-8 \sqrt{\Theta }\right)}  -4 \sqrt{\Theta } \sqrt{M \left(\sqrt{\pi } M-8 \sqrt{\Theta }\right)}\right)\right]\right\} \ln\left(\frac{|x|}{CD}\right)},
    \end{split}
\end{equation}
and
\begin{equation}
\begin{split}
\beta^{\Theta}_{\omega\omega'} &= i K e^{-i\omega^\prime v^{\Theta}_{0}}e^{-\pi \left\{   \frac{1}{{\pi ^{3/4} M-8 \sqrt[4]{\pi } \sqrt{\Theta }}}     \left[ \left(\pi ^{3/4} M^2-8 \sqrt[4]{\pi } \sqrt{\Theta } M+\sqrt{\pi } M \sqrt{M \left(\sqrt{\pi } M-8 \sqrt{\Theta }\right)}  -4 \sqrt{\Theta } \sqrt{M \left(\sqrt{\pi } M-8 \sqrt{\Theta }\right)}\right)\right] \right\} \omega} \\
& 
\int_{-\infty}^{0} \,\mathrm{d}x\,\left(\frac{\omega^\prime}{\omega}\right)^{1/2}e^{\omega^\prime x} \\
& \times e^{i\omega\left\{   \frac{1}{{\pi ^{3/4} M-8 \sqrt[4]{\pi } \sqrt{\Theta }}}     \left[2 \left(\pi ^{3/4} M^2-8 \sqrt[4]{\pi } \sqrt{\Theta } M+\sqrt{\pi } M \sqrt{M \left(\sqrt{\pi } M-8 \sqrt{\Theta }\right)}  -4 \sqrt{\Theta } \sqrt{M \left(\sqrt{\pi } M-8 \sqrt{\Theta }\right)}\right)\right] \right\} \ln\left(\frac{|x|}{CD}\right)},
\end{split}
\end{equation}
where $K$ is a normalization constant. This result indicates that the quantum amplitude governing particle production is influenced by non--commutative effects, with the parameter $\Theta$ introducing modifications to the metric structure, as argued previously.

Interestingly, despite the influence of quantum gravitational corrections on the quantum amplitude, the power spectrum at this stage still retains its blackbody nature. To confirm this behavior, it is necessary to evaluate the following expression:
\ie
    |\alpha^{\Theta}_{\omega\omega'}|^2 = e^{ \left\{  \frac{1}{{\pi ^{3/4} M-8 \sqrt[4]{\pi } \sqrt{\Theta }}}     \left[ 4 \left(\pi^{3/4} M^2-8 \sqrt[4]{\pi } \sqrt{\Theta } M+\sqrt{\pi } M \sqrt{M \left(\sqrt{\pi } M-8 \sqrt{\Theta }\right)}  -4 \sqrt{\Theta } \sqrt{M \left(\sqrt{\pi } M-8 \sqrt{\Theta }\right)}\right)\right]   \right\} \omega}|\beta^{\Theta}_{\omega\omega'}|^2\,.
\fe
Analyzing the flux of emitted particles within the frequency interval \(\omega\) to \(\omega + \mathrm{d}\omega\) \cite{o10}, the expression is obtained as
\ie
    \mathcal{P}(\omega, \Theta)=\frac{\mathrm{d}\omega}{2\pi}\frac{1}{\left \lvert\frac{\alpha^{\Theta}_{\omega\omega^\prime}}{\beta^{\Theta}_{\omega\omega^\prime}}\right \rvert^2-1}\, ,
\fe
or, therefore,
\ie
    \mathcal{P}(\omega, \Theta) = \frac{\mathrm{d}\omega}{2\pi}\frac{1}{e^{ \left\{  \frac{1}{{\pi ^{3/4} M-8 \sqrt[4]{\pi } \sqrt{\Theta }}}     \left[ 4 \left(\pi^{3/4} M^2-8 \sqrt[4]{\pi } \sqrt{\Theta } M+\sqrt{\pi } M \sqrt{M \left(\sqrt{\pi } M-8 \sqrt{\Theta }\right)}  -4 \sqrt{\Theta } \sqrt{M \left(\sqrt{\pi } M-8 \sqrt{\Theta }\right)}\right)\right]   \right\} \omega}-1}\,.
\fe

An important point to highlight is that when comparing the obtained expression with the Planck distribution, a direct correspondence can be observed, allowing for an interpretation of the radiation spectrum in terms of thermal emission
\ie
    \mathcal{P}(\omega, \Theta)=\frac{\mathrm{d}\omega}{2\pi}\frac{1}{e^{\frac{\omega}{T_{\Theta}}}-1},
\fe
so that
\ie
\begin{split}
\label{hawtemp1}
T_{\Theta} & = \frac{1}{\frac{4 \left(\pi ^{3/4} M^2-8 \sqrt[4]{\pi } \sqrt{\Theta } M+\sqrt{\pi } M \sqrt{M \left(\sqrt{\pi } M-8 \sqrt{\Theta }\right)}-4 \sqrt{\Theta } \sqrt{M \left(\sqrt{\pi } M-8 \sqrt{\Theta }\right)}\right)}{\pi ^{3/4} M-8 \sqrt[4]{\pi } \sqrt{\Theta }}} \\
& \approx \frac{1}{8 M} -\frac{\Theta }{2 \left(\pi  M^3\right)} -\frac{4 \Theta ^{3/2}}{\pi ^{3/2} M^4} - \frac{28 \Theta ^2}{\pi ^2 M^5}.
\end{split}
\fe

\begin{figure}
    \centering
      \includegraphics[scale=0.7]{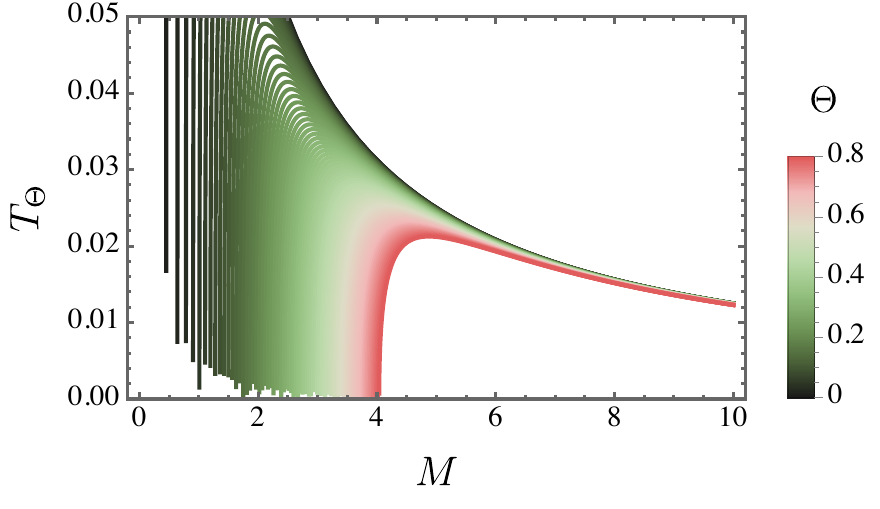}
    \caption{The Hawking temperature $T_{\Theta}$ expressed in terms of the black hole mass $M$ and the non--commutative parameter $\Theta$.}
    \label{fnfn}
\end{figure}

Above results is accomplished by considering the expansion up to the second order in $\Theta$. It reveals that the thermal spectrum remains consistent with the expected modifications \cite{Filho:2024zxx}. As discussed in the evaporation subsection, the Hawking temperature obtained in Eq. (\ref{hawtemp1}) matches the value derived from the surface gravity method in Eq. (\ref{haah}), confirming theoretical consistency. To further illustrate this behavior, Fig. \ref{fnfn} presents a graphical representation, including a comparison with the standard Schwarzschild scenario. In general, an increase in $\Theta$ leads to a smaller Hawking temperature.

In a complementaty manner, it is important to emphasize that energy conservation for the entire system has not yet been fully accounted for. As radiation is emitted, the black hole’s total mass diminishes, causing a progressive reduction in its size. To address this effect, the next section applies the tunneling approach formulated by Parikh and Wilczek \cite{011}. To implement this approach, we utilize the method outlined in \cite{011, vanzo2011tunnelling, parikh2004energy, calmet2023quantum}. Expressing the metric in the Painlevé--Gullstrand coordinates, it takes the form:
\ie
\mathrm{d}s^2 = - A_{\Theta}(r)\mathrm{d}t^2 + 2 H_{\Theta}(r) \mathrm{d}t \mathrm{d}r + \mathrm{d}r^2 + r^2\mathrm{d}\Omega^2,
\fe
where $H_{\Theta}(r) = \sqrt{A_{\Theta}(r)\big(B_{\Theta}(r)^{-1} - 1\big)}$. The probability of quantum tunneling is directly related to the imaginary component of the action \cite{parikh2004energy, vanzo2011tunnelling, calmet2023quantum}.

The trajectory of a freely moving particle in curved spacetime is governed by the action 
\ie
\mathcal{S}_{\Theta} = \int p^{\Theta}_\mu \, \mathrm{d}x^\mu.
\fe
To extract the imaginary contribution, we analyze the expression 
\ie 
p^{\Theta}_\mu \mathrm{d}x^\mu = p^{\Theta}_t \mathrm{d}t + p^{\Theta}_r \mathrm{d}r.
\fe
Since the term $p^{\Theta}_t \mathrm{d}t = -\omega \mathrm{d}t$ remains entirely real, it does not influence the imaginary part of the action. As a result, only the radial component contributes, leading to
\ie
\text{Im}\,\mathcal{S}_{\Theta} = \text{Im}\,\int_{r_i}^{r_f} \,p^{\Theta}_r\,\mathrm{d}r = \text{Im}\,\int_{r_i}^{r_f}\int_{0}^{p^{\Theta}_r} \,\mathrm{d}p^{\Theta '}_r\,\mathrm{d}r.
\fe

By applying Hamilton's equations to a system characterized by the Hamiltonian $H = M - \omega'$, it follows that $\mathrm{d}H = -\mathrm{d}\omega'$, where the energy of the emitted particle is constrained within the range $0 \leq \omega' \leq \omega$. This leads directly to the following expression:
\ie
\begin{split}
\text{Im}\, \mathcal{S}_{\Theta} & = \text{Im}\,\int_{r_i}^{r_f}\int_{M}^{M-\omega} \,\frac{\mathrm{d}H}{\mathrm{d}r/\mathrm{d}t}\,\mathrm{d} r  =\text{Im}\,\int_{r_i}^{r_f}\,\mathrm{d}r\int_{0}^{\omega} \,-\frac{\mathrm{d}\omega'}{\mathrm{d}r/\mathrm{d}t}\,.
\end{split}
\fe
Next, rearranging the integration order and introducing an appropriate substitution, the expression transforms into
\ie
    \frac{\mathrm{d}r}{\mathrm{d}t} = - H_{\Theta}(r)+\sqrt{A_{\Theta}(r) + H_{\Theta}(r)^2}=1-\sqrt{\frac{\Delta(r,\Theta,\omega^\prime)}{r}}, 
    \fe
where the function  
\ie
\Delta(r,\Theta,\omega^\prime) = \frac{(M - \omega^{\prime}) \left(\sqrt{\pi } r-4 \sqrt{\Theta }\right)}{r}
\fe  
is introduced to simplify the expression. Under this formulation, the result turns out to be
\ie
\label{ims}
\text{Im}\, \mathcal{S}_{\Theta} =\text{Im}\,\int_{0}^{\omega} -\mathrm{d}\omega'\int_{r_i}^{r_f}\,\frac{\mathrm{d}r}{1-\sqrt{\frac{\Delta(r,\,\Theta, \,\omega^\prime)}{r}}}.
\fe

Given that $\Theta$ represents a small parameter, the integrand in the previous expression can be expressed as
\ie
\begin{split}
\label{ims22}
\text{Im}\, \mathcal{S}_{\Theta} =  \text{Im}\,\int_{0}^{\omega} -\mathrm{d}\omega'\int_{r_i}^{r_f}\,\mathrm{d}r & \left\{ \frac{1}{1-\sqrt{2} \sqrt{\frac{M-\omega '}{r}}}  -\frac{2 \sqrt{\Theta } \left(\sqrt{\frac{2}{\pi }} \sqrt{\frac{M-\omega '}{r}}\right)}{r \left(\sqrt{2} \sqrt{\frac{M-\omega '}{r}}-1\right)^2} \right. \\
& \left. + \frac{2 \Theta  \left(\sqrt{2} r \sqrt{\frac{M-\omega '}{r}}-6 M+6 \omega '\right)}{\pi  r^3 \left(\sqrt{2} \sqrt{\frac{M-\omega '}{r}}-1\right)^3} \right\}.
\end{split}
\fe

An important observation is that replacing $M$ with $(M - \omega')$ introduces a new pole at $2(M - \omega')$ in the integral. Evaluating the contour integral around this pole in the counterclockwise direction leads to the following expression
\ie
    \text{Im}\, \mathcal{S}_{\Theta}  = 4\pi \omega \left( M - \frac{\omega}{2} \right) + 16 \Theta  \ln (M) - 16 \Theta  \ln (M-\omega ) .
\fe
As discussed in \cite{vanzo2011tunnelling}, the emission probability for the emitted particles, including the effects of non--commutative corrections, is given by:
\ie
\Gamma \sim e^{-2 \, \text{Im}\, S_{\Theta}} = e^{-8 \pi \omega \left( M - \frac{\omega}{2} \right) - 32 \Theta  \ln (M) + 32 \Theta  \ln (M-\omega)} .
\fe

In the limit $\omega \to 0$, the radiation spectrum recovers the standard Planckian distribution derived in Hawking's original formulation. Consequently, the emission spectrum reads
\ie
    \mathcal{P}(\omega,\Theta)=\frac{\mathrm{d}\omega}{2\pi}\frac{1}{e^{8 \pi \omega \left( M - \frac{\omega}{2} \right) + 32 \Theta  \ln (M) - 32 \Theta  \ln (M-\omega)
    }-1}.
\fe

The presence of $\omega$ and $\Theta$ introduces deviations in the emission spectrum from the standard blackbody distribution, a distinction that becomes clear upon analysis. When $\omega$ is small, the spectrum retains a Planck--like shape but with a modified Hawking temperature. Moreover, the number density of emitted particles can be reformulated using the tunneling probability as
\ie
n_{b}(\omega,\Theta) = \frac{\Gamma}{1 - \Gamma} = \frac{1}{e^{8 \pi \omega \left( M - \frac{\omega}{2} \right) + 32 \Theta  \ln (M) - 32 \Theta  \ln (M-\omega)} - 1}.
\fe

\begin{figure}
    \centering
      \includegraphics[scale=0.7]{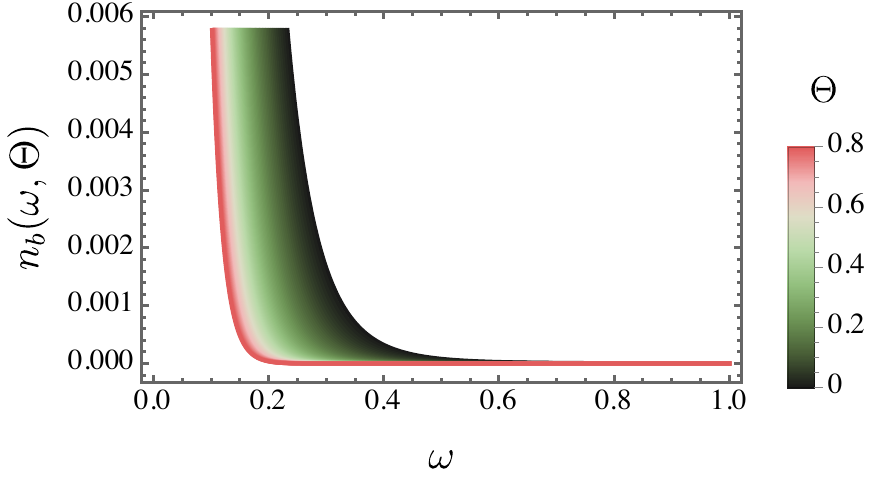}
    \caption{The particle density $n_{b}(\omega,\Theta)$ is plotted for various values of the non--commutative parameter $\Theta$.}
    \label{nbtheta}
\end{figure}

In summary, the Hawking amplitudes are modified by the presence of $\Theta$, leading to corrections in the power spectrum. These alterations cause the spectrum to deviate from the standard blackbody distribution, particularly when energy conservation is accounted for. To better understand the influence of $\Theta$ on $n_{b}(\omega,\Theta)$, Fig. \ref{nbtheta} depicts its variation with respect to the non--commutative parameter. The figure reveals that as $\Theta$ increases, the particle number density decreases. Moreover, in comparison to the non--commutative scenario, the Schwarzschild case corresponds to the uppermost curve.

%%%%%%%%%%%%%%%%%%%%%%%%%%%%%%%%%%%%%%%%%%%%%%%%%%%%%%%%%%%%%%%%%%%%%%%%%%%%%%%%%%%%%%%%%%%%%%%%%%%%%%%%%%%%%%%%%%%%%%%%%%%%%%%%%%%%%%%%%%%%%%%%%%%%%%%%%%%%%%%%%%%%%%%%%%%%%%%%%%%%%%%%%%%%%%%%%%%%%%%%%%%%%%%%%%%%%%%%%%%%%%%%%%%%%%%%%%%%%%%%%%%%%%%%%%%%%%%%%%%%%%%%

\subsection{Fermionic modes}

Black holes emit radiation at a characteristic temperature, producing a spectrum similar to blackbody radiation. However, this emission does not inherently account for modifications introduced by greybody factors. The radiation consists of particles with different spin states, including fermions. Research conducted by Kerner and Mann \cite{o69}, along with subsequent works \cite{o73,o72,o75,o70,o74,o71}, has established that both massless bosons and fermions radiate at the same temperature. Additionally, studies on spin--$1$ bosons have demonstrated that even with the inclusion of higher--order quantum corrections, the Hawking temperature remains unaffected \cite{o76,o77}.  

In the case of fermions, the action is often described through the phase of the spinor wave function, which is governed by the Hamilton--Jacobi equation. An alternative representation of the action can be formulated as \cite{o83, o84, vanzo2011tunnelling}:
\ie
\nonumber
\mathcal{S} = S^{(o)} + S^{(\psi_{\uparrow \downarrow})}.
\fe
In this formulation, $S^{(o)}$ corresponds to the classical action for scalar particles, while $S^{(\psi_{\uparrow \downarrow})}$ accounts for spin--dependent corrections. These corrections stem from the interaction between the particle’s intrinsic spin and the spin connection of the spacetime. However, they do not lead to singularities at the event horizon. Their overall effect remains minor, primarily influencing spin precession, allowing them to be safely disregarded in this analysis. Furthermore, the influence of emitted particles on the black hole’s angular momentum is negligible, especially for non-rotating black holes with masses far exceeding the Planck scale \cite{vanzo2011tunnelling}. Due to the statistical symmetry of particle emission, where spins are radiated in ``opposite directions'', the total angular momentum of the black hole remains effectively unchanged.

Building on the previous results, the focus now shifts to examining the tunneling process for fermionic particles as they traverse the event horizon of the black hole under consideration. The emission probability is evaluated within a Schwarzschild-like coordinate system, which is known to exhibit a coordinate singularity at the horizon. For alternative approaches that employ generalized Painlevé--Gullstrand and Kruskal--Szekeres coordinates, a detailed discussion can be found in \cite{o69}. To establish the framework for this analysis, we begin by introducing a general metric given by:
\ie
\mathrm{d}s^{2} = A_{\Theta}(r) \mathrm{d}t^{2} + \frac{1}{B_{\Theta}(r)}\mathrm{d}r^{2} + C_{\Theta}(r,\theta)\mathrm{d}\theta^{2} + D_{\Theta} (r,\theta) \mathrm{d}\varphi^{2}.
\fe

In the context of curved spacetime, the Dirac equation is expressed as:
\ie
\left(\gamma^\mu \nabla_\mu + \frac{m}{\hbar}\right) \psi = 0,
\fe
where
\ie
\nabla_\mu = \partial_\mu + \frac{\mathbbm{i}}{2} {\Gamma^\alpha_{\;\mu}}^{\;\beta} \,\Sigma_{\alpha\beta}
\fe
and 
\ie
\Sigma_{\alpha\beta} = \frac{\mathbbm{i}}{4} [\gamma_\alpha,  \gamma_\beta].
\fe

The $\gamma^\mu$ matrices obey the Clifford algebra, which is defined by the anticommutation relation  
\ie
\{\gamma_\alpha, \gamma_\beta\} = 2 g_{\alpha\beta} \mathbbm{1},
\fe  
with $\mathbbm{1}$ represents the $4 \times 4$ identity matrix. In this formulation, the $\gamma$ matrices are
\begin{eqnarray*}
 \gamma^{t} & = &\frac{\mathbbm{i}}{\sqrt{A_{\Theta}(r)}}\left( \begin{array}{cc}
\vec{1}& \vec{ 0} \\ 
\vec{ 0} & -\vec{ 1}%
\end{array}%
\right), \;\;
\gamma ^{r} =\sqrt{B_{\Theta}(r)}\left( 
\begin{array}{cc}
\vec{0} &  \vec{\sigma}_{3} \\ 
 \vec{\sigma}_{3} & \vec{0}%
\end{array}%
\right), \\
\gamma ^{\theta } &=&\frac{1}{\sqrt{C_{\Theta}(r)}}\left( 
\begin{array}{cc}
\vec{0} &  \vec{\sigma}_{1} \\ 
 \vec{\sigma}_{1} & \vec{0}%
\end{array}%
\right), \;\;
\gamma ^{\varphi} =\frac{1}{\sqrt{D_{\Theta}(r,\theta)} }\left( 
\begin{array}{cc}
\vec{0} &  \vec{\sigma}_{2} \\ 
 \vec{\sigma}_{2} & \vec{0}%
\end{array}%
\right),
\end{eqnarray*}%
in which $\vec{\sigma}$ represents the Pauli matrices, which satisfy the standard commutation relation:  
\ie
\sigma_i  \sigma_j = \mathbbm{1} \delta_{ij} + \mathbbm{i} \varepsilon_{ijk} \sigma_k, \quad \text{for} \quad i,j,k =1,2,3.
\fe

In contrast, the matrix associated with $\gamma^{5}$ is written
\ie
\gamma^{5} = \mathbbm{i} \gamma ^{t}\gamma ^{r}\gamma ^{\theta }\gamma ^{\varphi } = \mathbbm{i} \sqrt{\frac{B_{\Theta}(r)}{A_{\Theta}(r) \, C_{\Theta}(r,\theta) \, D_{\Theta}(r,\theta)}} \left( 
\begin{array}{cc}
\vec{ 0} & - \vec{ 1} \\ 
\vec{ 1} & \vec{ 0}%
\end{array}%
\right)\:.
\fe

To represent a Dirac field with spin directed along the positive $r$--axis, the following ansatz is chosen \cite{vagnozzi2022horizon}:
\ie
\psi_{\uparrow} = \left( \begin{array}{c}
\mathcal{H} \\ 
0 \\ 
\mathcal{Y} \\ 
0%
\end{array}%
\right)  e^{\frac{\mathbbm{i}}{\hbar }}\psi_{\uparrow}.
\label{spinupbh} 
\fe

The analysis is centered on the spin--up ($\uparrow$) configuration, while recognizing that the spin--down $(\downarrow$) case, aligned along the negative $r$--axis, follows an analogous procedure. By inserting the ansatz (\ref{spinupbh}) into the Dirac equation, the following expression is obtained:
\ie
\begin{split}
-\left( \frac{\mathbbm{i} \,\mathcal{H}}{\sqrt{A_{\Theta}(r)}}\,\partial _{t} \psi_{\uparrow} + \mathcal{Y} \sqrt{B_{\Theta}(r)} \,\partial_{r} \psi_{\uparrow}\right) + m \, \mathcal{H}  &=0, \\
-\frac{\mathcal{Y}}{r}\left( \partial _{\theta }\psi_{\uparrow} +\frac{\mathbbm{i}}{\sin \theta } \, \partial _{\varphi }\psi_{\uparrow}\right) &= 0, \\
\left( \frac{\mathbbm{i} \,\mathcal{Y}}{\sqrt{A_{\Theta}(r)}}\,\partial _{t}\psi_{\uparrow} - \mathcal{H} \sqrt{B_{\Theta}(r)}\,\partial_{r}\psi_{\uparrow}\right) + m \, \mathcal{Y} &= 0, \\
-\frac{\mathcal{H}}{r}\left(\partial _{\theta }\psi_{\uparrow} + \frac{\mathbbm{i}}{\sin \theta }\,\partial _{\varphi }\psi_{\uparrow}\right) &= 0.
\end{split}
\fe%
Considering the leading order in $\hbar$, the action reads 
\ie
\psi_{\uparrow}=- \omega\, t + \xi(r) + L(\theta ,\varphi )
\fe
which leads to the following equations \cite{vanzo2011tunnelling}:
\begin{eqnarray}
\left( \frac{\mathbbm{i}\, \omega\, \mathcal{H}}{\sqrt{A_{\Theta}(r)}} - \mathcal{Y} \sqrt{B_{\Theta}(r)}\,  \xi^{\prime }(r)\right) +m\, {\mathbbm{i}}\mathcal{H} &=&0,
\label{bhspin5} \\
-\frac{{\mathcal{Y}}}{r}\left( L_{\theta }+\frac{\mathbbm{i}}{\sin \theta }L_{\varphi }\right) &=&0,
\label{bhspin6} \\
-\left( \frac{\mathbbm{i}\,\omega\, \mathcal{Y}}{\sqrt{A_{\Theta}(r)}} + \mathcal{H}\sqrt{B_{\Theta}(r)}\,  \xi^{\prime }(r)\right) + m \, {\mathbbm{i}} \mathcal{Y} &=&0,
\label{bhspin7} \\
-\frac{\mathcal{H}}{r}\left( L_{\theta } + \frac{\mathbbm{i}}{\sin \theta }L_{\varphi }\right) &=& 0.
\label{bhspin8}
\end{eqnarray}

The expressions for $\mathcal{H}$ and $\mathcal{Y}$ do not alter the conclusion that Eqs. (\ref{bhspin6}) and (\ref{bhspin8}) impose the constraint $L_{\theta} + \mathbbm{i}(\sin \theta)^{-1} L_{\varphi} = 0$. This implies that $L(\theta, \varphi)$ must necessarily be complex, a condition that holds for both outgoing and ingoing solutions. As a result, when computing the ratio between outgoing and ingoing probabilities, the terms involving $L$ cancel out, allowing it to be disregarded in the subsequent analysis. For massless particles, Eqs. (\ref{bhspin5}) and (\ref{bhspin7}) lead to two distinct possible solutions:
\ie
\mathcal{H} = -\mathbbm{i} \mathcal{Y}, \qquad \xi^{\prime }(r) = \xi_{\text{out}}' = \frac{\omega}{\sqrt{A_{\Theta}(r)B_{\Theta}(r)}},
\fe
\ie
\mathcal{H} = \mathbbm{i} \mathcal{Y}, \qquad \xi^{\prime }(r) = \xi_{\text{in}}' = - \frac{\omega}{\sqrt{A_{\Theta}(r)B_{\Theta}(r)}},
\fe
where $\xi_{\text{out}}$ and $\xi_{\text{in}}$ represent the solutions associated with outgoing and ingoing particles, respectively \cite{vanzo2011tunnelling}. This leads to the expression for the total tunneling probability, given by  
\ie
\Gamma_{\psi} \sim e^{-2 \, \text{Im} \, (\xi_{\text{out}} - \xi_{\text{in}})}.
\fe  
Accordingly, the result can be written as:
\ie
\mathcal \xi_{ \text{out}}(r)= -  \xi_{ \text{in}} (r) = \int \mathrm{d} r \,\frac{\omega}{\sqrt{A_{\Theta}(r)B_{\Theta}(r)}}\:.
\fe

It is important to highlight that, under the dominant energy condition and the Einstein field equations, the functions $A_{\Theta}(r)$ and $B_{\Theta}(r)$ possess the same roots. Consequently, in the vicinity of $r = r_{+}$, their behavior can be approximated to first order as:
\ie
A_{\Theta}(r) B_{\Theta}(r) = A_{\Theta}'(r_{+})B_{\Theta}'(r_{+})(r - r_{+})^2 + \dots
\fe
which reveals the existence of a simple pole with a well--defined coefficient. By applying Feynman’s method, the following expression is therefore
\ie
2\mbox{Im}\;\left(\xi_{\text{out}} -  \xi_{\text{in}} \right) =\mbox{Im}\int \mathrm{d} r \,\frac{4\omega}{\sqrt{A_{\Theta}(r)B_{\Theta}(r)}}=\frac{2\pi\omega}{\kappa}.
\fe
In this framework, the surface gravity is expressed as  
\ie
\kappa = \frac{1}{2} \sqrt{A_{\Theta}'(r_{+}) B_{\Theta}'(r_{+})}.
\fe  
Given these conditions, the tunneling probability follows the relation  
\ie
\Gamma_{\psi} \sim e^{-\frac{2 \pi \omega}{\kappa}}.
\fe  
This expression determines the particle number density $n_{f}(\omega,\Theta)$ corresponding to the black hole in question below
\ie
n_{f}(\omega,\Theta) = \frac{\Gamma_{\psi}}{1+\Gamma_{\psi}}  = \frac{1}{\exp \left(\frac{\sqrt{2} \pi ^{3/4} \omega  \left(\sqrt{M \left(\sqrt{\pi } M-8 \sqrt{\Theta }\right)}+\sqrt[4]{\pi } M\right)^3}{M \sqrt{\left(\sqrt{\pi } M-8 \sqrt{\Theta }\right) \left(-4 \sqrt{\Theta }+\sqrt[4]{\pi } \sqrt{M \left(\sqrt{\pi } M-8 \sqrt{\Theta }\right)}+\sqrt{\pi } M\right)}}\right)+1}.
\fe

\begin{figure}
    \centering
      \includegraphics[scale=0.7]{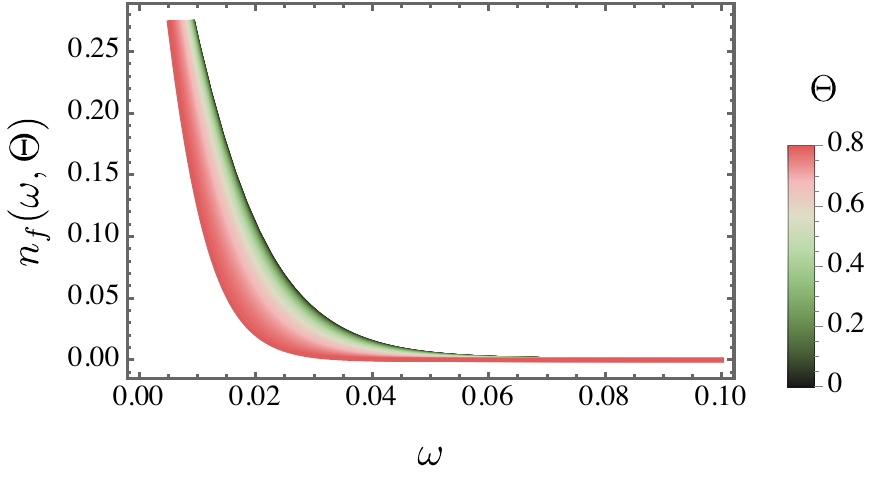}
    \caption{The particle density $n_{f}(\omega,\Theta)$ is plotted for different values of the non--commutative parameter $\Theta$.}
    \label{partfermions}
\end{figure}

In Fig. \ref{partfermions}, the dependence of $n_{f}(\omega,\Theta)$ on the non--commutative parameter $\Theta$ is illustrated, with a comparison to the Schwarzschild limit. As in the bosonic case, increasing $\Theta$ leads to a reduction in fermionic particle density. The Schwarzschild scenario, where $\Theta = 0$, stands out as the uppermost curve in the plot.

%%%%%%%%%%%%%%%%%%%%%%%%%%%%%%%%%%%%%%%%%%%%%%%%%%%%%%%%%%%%%%%%%%%%%%%%%%%%%%%%%%%%%%%%%%%%%%%%%%%%%%%%%%%%%%%%%%%%%%%%%%%%%%%%%%%%%%%%%%%%%%%%%%%%%%%%%%%%%%%%%%%%%%%%%%%%%%%%%%%%%%%%%%%%%%%%%%%%%%%%%%%%%%%%%%%%%%%%%%%%%%%%%%%%%%%%%%%%%%%%%%%%%%%%%%%%%%%%%%%%%%%%%%%%%%%%%%%%%%%%%%%%%%%%%%%%%%%%%%%%%%%%%%%%%%%%%%%%%%%%%%%%%%%%%%%%%%%%%%%%%%%%%%%%%%%%%%%%%%%%%%%%%%%%%%%%%%%%%%%%%%%%%%%%%%%%%%%%%%%%%%%%%%%%%%%%%%%%%%%%

\section{The black hole evaporation}

An essential factor to analyze in this context is the duration of the black hole’s evaporation process. This can be determined as \cite{23araujo2023thermodynamics,ong2018effective}
\ie
\label{evava}
\frac{\mathrm{d}M}{\mathrm{d}\tau} = - \alpha \sigma a T_{\Theta}^{4}.
\fe
In this formulation, $a$ corresponds to the radiation constant, $\sigma$ represents the cross--sectional area, and $\alpha$ accounts for the greybody factor. Unlike the approach used in earlier sections to determine the Hawking temperature $T_{\Theta}$, the analysis here adopts the surface gravity method to explore the black hole’s thermal properties.

The radiation emitted by the black hole is predominantly composed of massless particles, such as photons and neutrinos. Due to the challenges associated with the emission of long--wavelength particles, the high--frequency limit is often adopted. In this regime, radiation propagation is effectively described by null geodesics, a principle known as the geometrical optics approximation \cite{hiscock1990evolution,page1976particle}. Consequently, the cross--sectional area $\sigma$ is expressed as $\pi \mathcal{R}^2$, where $\mathcal{R}$ corresponds to the shadow radius. Additionally, under this framework, the greybody factor approaches $\alpha \to 1$ \cite{liang2025einstein}.

In addition, the following expression holds:
\ie 
\label{surface3}
\kappa = {\left.\frac{A_{\Theta}^{\prime}(r)}{2} \right|_{r = {r_{+}}}}.
\fe

Hawking’s groundbreaking discovery, detailed in Ref. \cite{hawking1975particle}, demonstrated that black holes emit radiation, leading to the definition of the Hawking temperature. This temperature is expressed as $T_{\Theta} = \kappa / 2\pi$, and within the framework of our analysis, it takes the following form:
\ie
\label{haah}
T_{\Theta}   \approx \frac{1}{8 M} -\frac{\Theta }{2 \left(\pi  M^3\right)} -\frac{4 \Theta ^{3/2}}{\pi ^{3/2} M^4} - \frac{28 \Theta ^2}{\pi ^2 M^5}.
\fe

Interestingly, the Hawking temperature derived via the surface gravity method coincides with the result obtained from the analysis of Hawking radiation using Bogoliubov coefficients, confirming the consistency of both approaches. With this validation in place, attention now turns to the evaporation dynamics of the non--commutative black hole. A crucial aspect of this process is determining the remnant mass, $M_{\text{rem}}$. In this scenario, as the black hole reaches the final stage of evaporation, where $T_{\Theta} \to 0$, the mass satisfies the following relation:
\ie
\label{regnnmavnatsmass}
M_{rem} = \frac{2 \sqrt{\Theta }}{\sqrt{\pi }}.
\fe
From the given equation, it becomes evident that the only parameter affecting the modification of $M_{\text{rem}}$ is $\Theta$. It is worth noting that in deriving the expression for $M_{\text{rem}}$, only the first two terms of the Hawking temperature expansion from Eq. (\ref{haah}) were considered.

Additionally, within the framework of the geometric optics approximation, $\sigma$ is interpreted as the effective cross--section for photon capture:
\ie
\begin{split}
& \sigma =  \pi \left. \left( \frac{D_{\Theta}(r,\theta)}{A_{\Theta}(r)} \right)  \right|_{r = {r_{ph}}} \\
& = \frac{\pi\left(\sqrt{M \left(9 \sqrt{\pi } M-64 \sqrt{\Theta }\right)}+3 \sqrt[4]{\pi } M\right)^4}{8 \left(3 \pi  M^2-16 \sqrt{\pi } \sqrt{\Theta } M+\pi ^{3/4} M \sqrt{M \left(9 \sqrt{\pi } M-64 \sqrt{\Theta }\right)}\right)}.
\end{split}
\fe
Here, $r_{ph}$ represents the radius of the photon sphere, defined as follows. It is worth mentioning that a recent study \cite{Filho:2024zxx} also derived the photon sphere radius. However, a typographical error was present in its expression, which has been corrected in this work
\ie
r_{ph} = \frac{1}{2} \left( 3 M + \frac{\sqrt{M \left(9 \sqrt{\pi } M-64 \sqrt{\Theta }\right)}}{\sqrt[4]{\pi }}\right).
\fe

It is important to note that in the limit $\Theta \to 0$, both $\sigma$ and $r_{ph}$ revert to their standard Schwarzschild values, specifically $27\pi M^2$ and $3M$, respectively, as expected.  Using these results, we substitute the obtained expressions into Eq. (\ref{evava}), leading to:
\ie
\begin{split}
\frac{\mathrm{d}M}{\mathrm{d}\tau} & = 
-\frac{\Upsilon \left(\sqrt{M \left(9 \sqrt{\pi } M-64 \sqrt{\Theta }\right)}+3 \sqrt[4]{\pi } M\right)^4 \left(\pi  M^2-4 \Theta \right)^4}{32768 \pi ^{15/2} M^{13} \left(-16 \sqrt{\Theta }+\sqrt[4]{\pi } \sqrt{M \left(9 \sqrt{\pi } M-64 \sqrt{\Theta }\right)}+3 \sqrt{\pi } M\right)},
\end{split}
\fe
in which $\Upsilon = a \alpha$. With this definition, the problem reduces to solving the following equation \cite{Filho:2024zxx}:
\ie
\begin{split}
& \int_{0}^{t_{evap}} \Upsilon \mathrm{d}\tau  = \\
& \int_{M_{i}}^{M_{rem}}
\mathrm{d}M\left[-\frac{ \left(\sqrt{M \left(9 \sqrt{\pi } M-64 \sqrt{\Theta }\right)}+3 \sqrt[4]{\pi } M\right)^4 \left(\pi  M^2-4 \Theta \right)^4}{32768 \pi ^{15/2} M^{13} \left(-16 \sqrt{\Theta }+\sqrt[4]{\pi } \sqrt{M \left(9 \sqrt{\pi } M-64 \sqrt{\Theta }\right)}+3 \sqrt{\pi } M\right)}\right]^{-1} ,
\end{split}
\fe
where $M_{i}$ represents the initial mass of the black hole, and $t_{\text{evap}}$ denotes the time required for the black hole to reach the final stage of its evaporation. By solving the equation and considering the first--order expansion in the limit $\Theta \ll 1$, the general expression for $t_{\text{evap}}$ is obtained as
\ie
\begin{split}
& t_{evap}  = -\frac{4096}{81}  \, \pi ^3 (M^{3}_{f} - M^{3}_{i}) - \frac{16384}{81} \sqrt{\Theta } \,\pi ^{5/2} (M^2_{f} - M^2_{i}) - \frac{2818048}{729} \Theta \, \pi ^2 (M_{f} - M_{i}),
\end{split}
\fe
where, for simplicity, we have assumed $\Upsilon \approx 1$. Additionally, when the black hole reaches its final stage, characterized by $M_{f} = M_{\text{rem}} \to 2\sqrt{\Theta}/\sqrt{\pi}$, the expression simplifies to:
\ie
t_{evapfinal} = \frac{4096}{729} \pi ^{3/2} \left(\sqrt{\pi } M_{i} - 2 \sqrt{\Theta }\right) \left(796 \,\Theta +9 \pi  M_{i}^2 + 54 \sqrt{\pi } \sqrt{\Theta } M_{i} \right).
\fe

\begin{figure}
    \centering
      \includegraphics[scale=0.7]{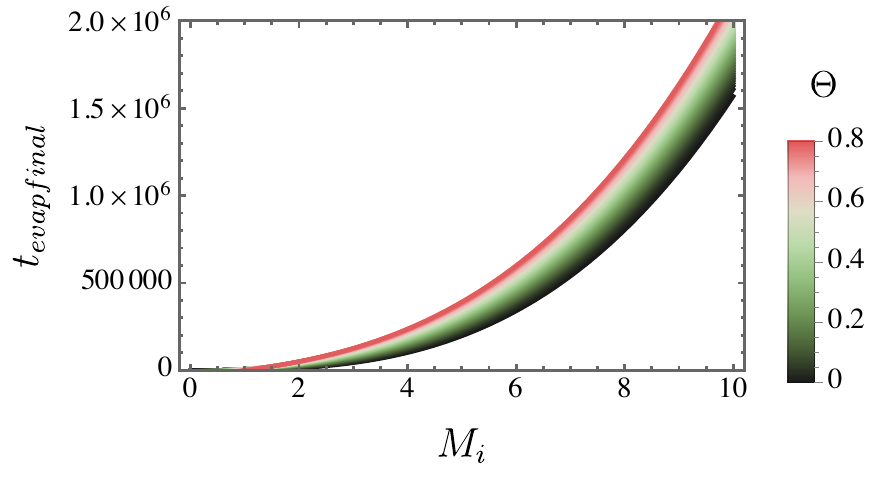}
    \caption{The final evaporation time $t_{\text{evapfinal}}$ is presented for various values of the non--commutative parameter $\Theta$ and initial mass $M_{i}$.}
    \label{evap}
\end{figure}

Fig. \ref{evap} illustrates the final evaporation time, $t_{\text{evapfinal}}$, for various values of $\Theta$ and initial mass $M_{i}$. The plot reveals that as $\Theta$ increases, $t_{\text{evapfinal}}$ decreases. In other words, a larger $\Theta$ causes the evaporation process to proceed at a slower rate.

%%%%%%%%%%%%%%%%%%%%%%%%%%%%%%%%%%%%%%%%%%%%%%%%%%%%%%%%%%%%%%%%%%%%%%%%%%%%%%%%%%%%%%%%%%%%%%%%%%%%%%%%%%%%%%%%%%%%%%%%%%%%%%%%%%%%%%%%%%%%%%%%%%%%%%%%%%%%%%%%%%%%%%%%%%%%%%%%%%%%%%%%%%%%%%%%%%%%%%%%%%%%%%%%%%%%%%%%%%%%%%%%%%%%%%%%%%%%%%%%%%%%%%%%%%%%%%%%%%%%%%%%%%%%%%%%%%%%%%%%%%%%%%%%%%%%%%%%%%%%%%%%%%%%%%%%%%%%%%%%%%%%%%%%%%%%%%%%%%%%%%%%%%%%%%%%%%%%%%%%%%%%%%%%%%%%%%%%%%%%%%%%%%%%%%%%%%%%%%%%%%%%%%%%%%%%%%%%%%%%

\section{Greybody factors: bosonic case}\label{Tbound}

It is well--known that black holes emit radiation. Such an emission is recognized in the literature as Hawking radiation. Fundamentally, it is as a consequence of quantum effects occurring near the event horizon. Once emitted, this radiation propagates outward but is influenced by the curvature of spacetime surrounding the black hole, leading to modifications in both its spectral distribution and intensity before reaching a distant observer. As a result, an observer at infinity perceives a spectrum that deviates from an ideal blackbody distribution.  

The extent of this deviation is quantified by the greybody factor, which characterizes how the black hole’s gravitational potential alters the radiation's propagation. In this section, we analyze the greybody factor for massless spin--$0$ and spin--$1/2$ particles emitted from it. The study is conducted using general semi--analytic methods \cite{sakalli2022topical,boonserm2019greybody,ovgun2024shadow,al2024fermionic,boonserm2008transmission,heidari2025absorption,heidari2024scattering}. The upper bound on the greybody factor, denoted as $T_b$, is expressed as:
\ie
\label{Tb0}
{T_b} \ge {\mathop{\rm sech}\nolimits} ^2 \left(\int_\infty^ {+\infty} {G(\omega,\Theta) \rm{d}}r^{*}\right),
\fe
with 
\ie
\label{Tb00}
G(\omega,\Theta) = \frac{{\sqrt {{{(h')}^2} + {{({\omega ^2} - {V^{(\text{s,v,t})}_{\text{eff}}} - {h^2})}^2}} }}{{2h}}.
\fe

In this context, $h$ represents a positive function that meets the conditions $h(+\infty) = h(-\infty) = \omega$. Meanwhile, $V^{(\text{s,v,t})}_{\text{eff}}$ denotes the effective potential, where the superscripts ``\text{s}'', ``\text{v}'', and ``\text{t}'' refer to scalar, vector, and tensor perturbations, respectively. When $h$ is set equal to $\omega$, we obtain the following expression:
\ie
\label{greybody}
T_{b}^{(\text{s,v,t})} \ge {\mathop{\rm sech}\nolimits} ^2 \left(\int_{-\infty}^ {+\infty} \frac{V^{(\text{s,v,t})}_{\text{eff}}} {2\omega}\mathrm{d}r^{*}\right)={\mathop{\rm sech}\nolimits} ^2 \left(\int_{r_{ h}}^ {+\infty} \frac{V^{(\text{s,v,t})}_{\text{eff}}} {2\omega\sqrt{A_{\Theta}(r)B_{\Theta}(r)}}\mathrm{d}r\right).
\fe

As previously discussed, calculating the greybody factors requires determining the corresponding effective potential, which varies depending on the type of perturbation being considered. These perturbations are classified as scalar ($V^{(\text{s})}_{\text{eff}}$), vector ($V^{(\text{v})}_{\text{eff}}$), or tensor ($V^{(\text{t})}_{\text{eff}}$). In the following, we will derive each of these potentials to compute the greybody factors. Additionally, by utilizing these effective potentials, it becomes possible to study the evolution of the perturbations over time as well.

%%%%%%%%%%%%%%%%%%%%%%%%%%%%%%%%%%%%%%%%%%%%%%%%%%%%%%%%%%%%%%%%%%%%%%%%%%%%%%%%%%%%%%%%%%%%%%%%%%%%%%%%%%%%%%%%%%%%%%%%%%%%%%%%%%%%%%%%%%%%%%%%%%%%%%%%%%%%%%%%%%%%%%%%%%%%%%%%%%%%%%%%%%%%%%%%%%%%%%%%%%%%%%%%%%%%%%%%%%%%%%%%%%%%%%%%%%%%%%%%%%%%%%%%%%%%%%%%%%%%%%%%%%%%%%%%%%%%%%%%%%%%%%%%%%%%%%%%%%%%%%%%

\subsection{Scalar perturbations}

To begin, let us consider the Klein–-Gordon equation, which governs the behavior of scalar fields in a curved spacetime
\ie
\nonumber
\frac{1}{\sqrt{-g_{\Theta}}}\partial_{\mu}[g_{\Theta}^{\mu\nu}\sqrt{-g_{\Theta}}\partial_{\nu}\Phi(t, r, \theta, \varphi)] = 0.
\fe
In this context, we focus on examining the scalar field as a small perturbation. Therefore, it takes the following form:
\ie
\label{1j2e3k4j5sk}
\begin{split}
-& \frac{1}{A_{\Theta}(r)} \frac{\partial^{2} \Phi (t, r, \theta, \varphi)}{\partial t^{2}} + \frac{1}{r^{2}} \left[  \frac{\partial}{\partial r} \left(  A_{\Theta}(r) \, r^{2}  \frac{\partial \Phi(t, r, \theta, \varphi)}{\partial r}  \right)  \right] \\  + & \frac{1}{r^{2} \sin \theta}  \left[  \frac{\partial }{\partial \theta} \left( \sin \theta \frac{\partial}{\partial \theta} \Phi(t, r, \theta, \varphi)   \right)        \right] 
 +  \frac{1}{r^{2} \sin^{2}}  \frac{\partial^{2} \Phi(t, r, \theta, \varphi)}{\partial \phi^{2}} = 0.
\end{split}
\fe
Here, $\sqrt{-g_{\Theta}} = r^{2}\sin\theta$, where the determinant of the metric is considered. Given the spherical symmetry of the system, the scalar field can be decomposed into spherical harmonics, expressed as 
\ie
\Phi(t, r, \theta, \varphi) = \sum_{l=0}^{\infty} \sum_{m=-l}^{l}  Y_{lm}(\theta, \varphi) \frac{\Psi(t,r)}{r} ,
\fe
with $Y_{lm}(\theta, \varphi)$ stands for the spherical harmonics. Taking into account such a decomposition, the radial equation in Eq. (\ref{1j2e3k4j5sk}) reads
\ie
\frac{\partial^{2}\Psi(t,r)}{\partial t^{2}}  + \frac{A_{\Theta}(r)}{r} \left\{ \frac{\partial }{\partial r}  \left[ A_{\Theta}(r) r^{2} \frac{\partial}{\partial r}  \left( \frac{\Psi(t,r)}{r} \right)   \right]     \right\} - A_{\Theta}(r) \frac{\ell(\ell + 1)}{r^{2}}\Psi(t,r) = 0 .
\fe

The Klein--Gordon equation can be recast into a Schrödinger--like form by introducing the tortoise coordinate, as defined in Eq. (\ref{1t2o3r4t5o6ise})
\ie
-\frac{\partial^2 \Psi(t,r)}{\partial t^2} + \frac{\partial^2 \Psi(t,r)}{\partial r^{*2}} + V^{(\text{s})}_{\text{eff}}(r) \Psi(t,r) = 0,
\fe 
with the effective potential, being given by
\ie
\label{Vscalar}
\begin{split}
V^{(\text{s})}_{\text{eff}}(r,\Theta) = A_{\Theta}(r) \left[ \frac{\ell (\ell+1)}{r^2} + \frac{2 M}{r^3}-\frac{16 \sqrt{\Theta } M}{\sqrt{\pi } r^4} \right]
\end{split}.
\fe

Fig. \ref{vs} displays the variation of $V^{(\text{s})}_{\text{eff}}(r, \Theta)$ for different values of $\Theta$ and $\ell$. In general, as $\Theta$ increases, the effective potential $V^{(\text{s})}_{\text{eff}}(r, \Theta)$ decreases for $\ell = 0$. However, for higher angular momentum values, specifically $\ell = 1$ and $\ell = 2$ (with $r \gtrsim 1.89$), an increase in $\Theta$ results in a corresponding rise in the effective potential.

\begin{figure}
    \centering
      \includegraphics[scale=0.519]{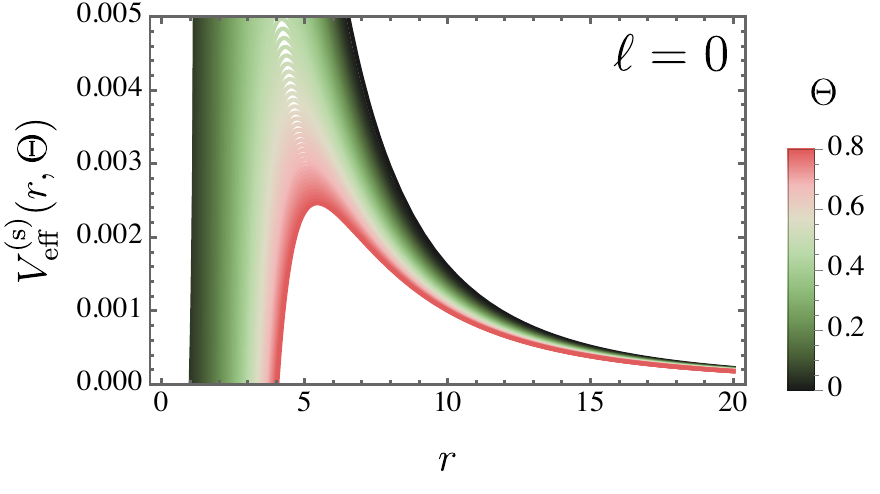}
      \includegraphics[scale=0.5]{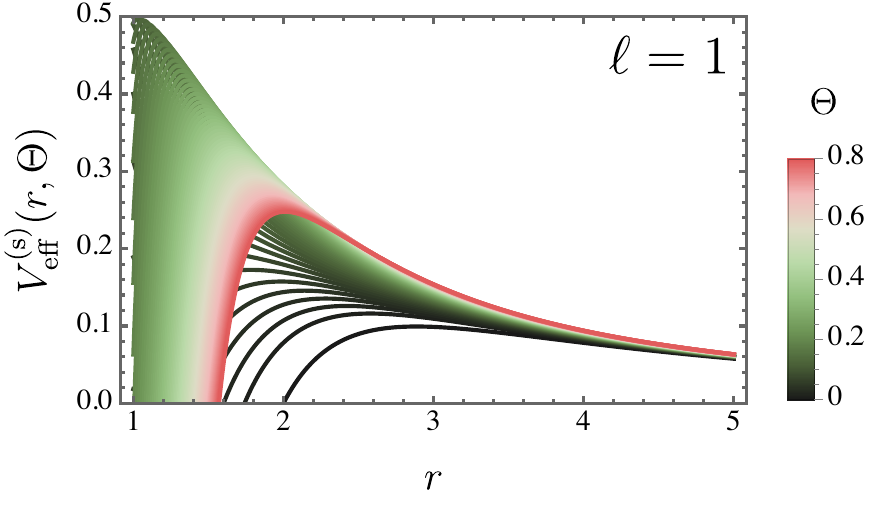}
      \includegraphics[scale=0.5]{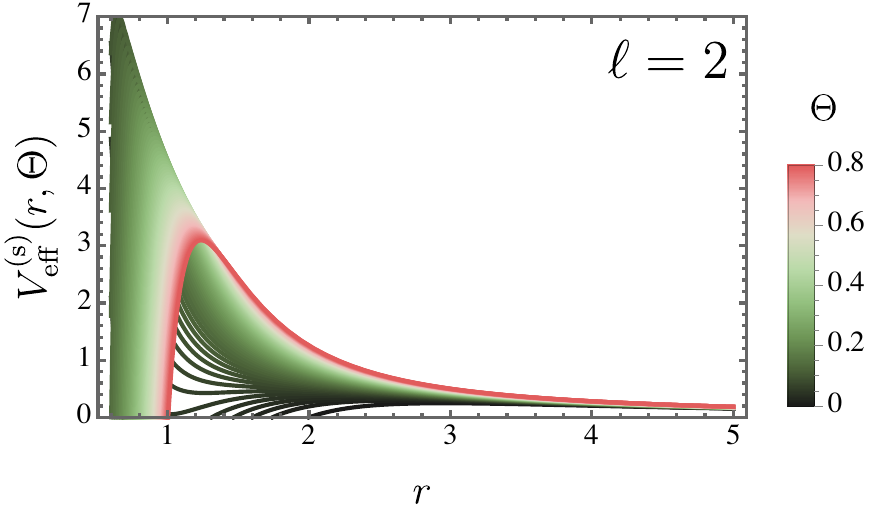}
    \caption{The scalar effective potential, $V^{(\text{s})}_{\text{eff}}(r, \Theta)$, is presented for various values of $\Theta$ and $\ell$.}
    \label{vs}
\end{figure}

\begin{figure}
    \centering
      \includegraphics[scale=0.54]{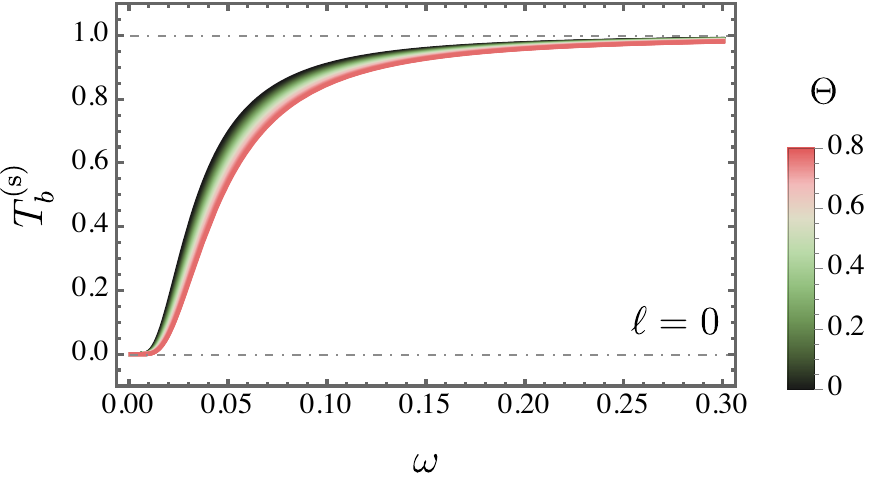}
      \includegraphics[scale=0.54]{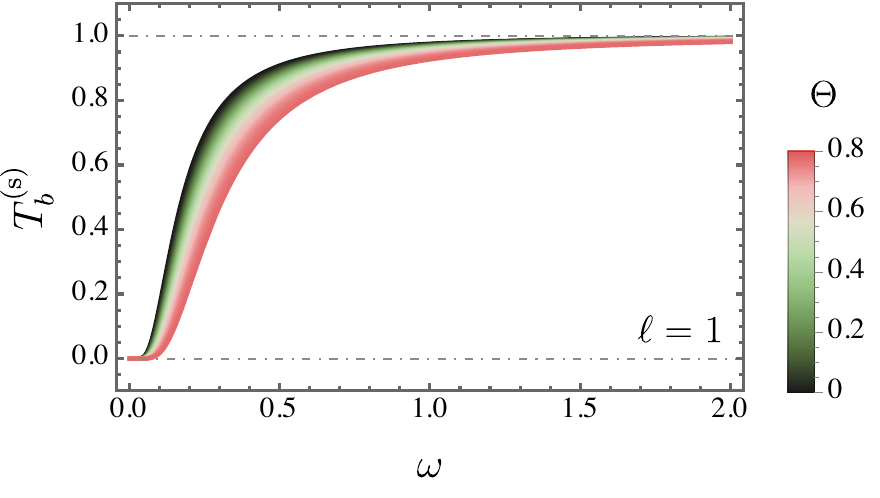}
      \includegraphics[scale=0.54]{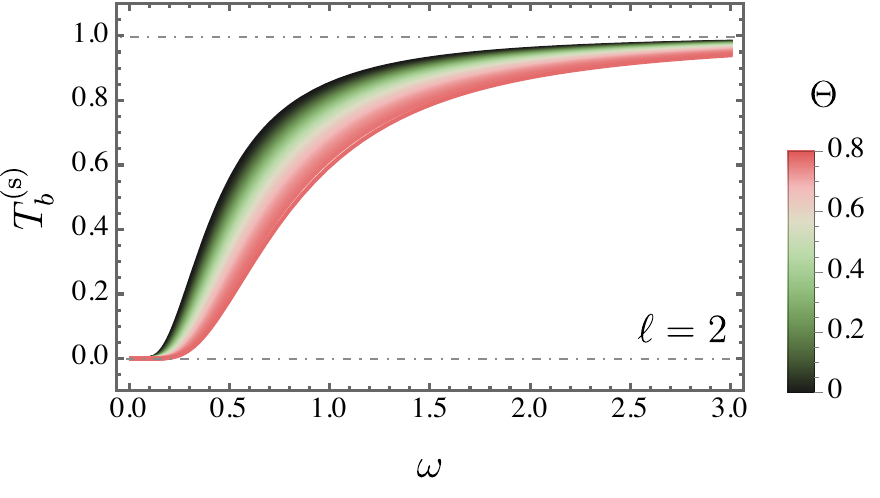}
    \caption{The greybody factors for the scalar perturbations $T_{b}^{(\text{s})}$ are shown for different values of $\Theta$ and $\ell$.}
    \label{tbs}
\end{figure}

Substituting Eq. (\ref{Vscalar}) into Eq. (\ref{greybody}), the following expression is obtained:
\ie
\begin{split}
T_b^{\text{(s)}} = & \, \text{sech}^{2} \left(\int_{r_{ h}}^ {+\infty} \frac{V^{(\text{s})}_{\text{eff}}} {2\omega\sqrt{A_{\Theta}(r)B_{\Theta}(r)}}\mathrm{d}r\right) \\
= & \, \text{sech}^{2} \left\{ {\frac{1}{2\omega}\times}  \frac{\sqrt[4]{\pi } M}{3 \left(\sqrt{M \left(\sqrt{\pi } M-8 \sqrt{\Theta }\right)}+\sqrt[4]{\pi } M\right)^3} \times \left[ 3 \sqrt[4]{\pi } (2 \ell (\ell+1)+1) \sqrt{M \left(\sqrt{\pi } M-8 \sqrt{\Theta }\right)}  \right.\right. \\
& \left.\left. +3 \sqrt{\pi } (2 \ell (\ell+1)+1) M  - 8 \sqrt{\Theta } (3 l (l+1)+2)  \right]  \right\}.
\end{split}
\fe
Fig. \ref{tbs} illustrates the greybody factors for scalar perturbations, $T_b^{\text{(s)}}$, across various values of $\Theta$ and $\ell$. For the values of $\ell$ considered in this analysis, the non--commutative parameter $\Theta$ leads to a reduction in the magnitude of the greybody factors.

%%%%%%%%%%%%%%%%%%%%%%%%%%%%%%%%%%%%%%%%%%%%%%%%%%%%%%%%%%%%%%%%%%%%%%%%%%%%%%%%%%%%%%%%%%%%%%%%%%%%%%%%%%%%%%%%%%%%%%%%%%%%%%%%%%%%%%%%%%%%%%%%%%%%%%%%%%%%%%%%%%%%%%%%%%%%%%%%%%%%%%%%%%%%%%%%%%%%%%%%%%%%%%%%%%%%%%%%%%%%%%%%%%%%%%%%%%%%%%%%%%%%%%%%%%%%%%%%%%%%%%%%%%%%%%%%%%%%%%%%%%%%%%%%%%%%%%%%%%%%%%%%

\subsection{Vectorial perturbations}

We now focus on the electromagnetic perturbation, which requires adopting the standard tetrad formalism \cite{chandrasekhar1998mathematical,Bouhmadi-Lopez:2020oia,Gogoi:2023kjt}. In this formalism, a basis $e_\mu^{a}$ is constructed with respect to the black hole metric $g^{\Theta}_{\mu\nu}$. These ones must satisfy
\ie
e^{a}_\mu e^\mu_{b} = \delta^{a}_{b}, \, \, \, \,
e^{a}_\mu e^\nu_{a} = \delta^{\nu}_{\mu}, \, \, \, \,
e^{a}_\mu = g^{\Theta}_{\mu\nu} \eta_{\Theta}^{a b} e^\nu_{b}, \, \, \, \,
g^{\Theta}_{\mu\nu} = \eta^{\Theta}_{a b} e^{a}_\mu e^{b}_\nu = e_{a\mu} e^{a}_\nu.
\fe

Within the tetrad formalism for electromagnetic perturbations, the Bianchi identity for the field strength, $\mathfrak{F}_{[ab|c]} = 0$, implies the following relation
\begin{align}
\left( r \sqrt{A_{\Theta}(r)}\, \mathfrak{F}_{t \phi}\right)_{,r} + r \sqrt{B_{\Theta}(r)}\,
\mathfrak{F}_{\phi r, t} &=0,  \label{1e2d3e4m1} \\
\left( r \sqrt{A_{\Theta}(r)}\, \mathfrak{F}_{ t \phi}\sin\theta\right)_{,\theta} + r^2
\sin\theta\, \mathfrak{F}_{\phi r, t} &=0.  \label{1e2d3e4m2}
\end{align}
As a result, we have
\ie
\eta_{\Theta}^{b c}\! \left( \mathfrak{F}_{a b} \right)_{|c} =0.
\fe

Above expression can also be rewritten in terms of spherical polar coordinates as
\ie  \label{1e2d3e4m3}
\left( r \sqrt{A_{\Theta}(r)}\, \mathfrak{F}_{\phi r}\right)_{,r} + \sqrt{A_{\Theta}(r) B_{\Theta}(r)}%
\, \mathfrak{F}_{\phi \theta,\theta} + r \sqrt{B_{\Theta}(r)}\, \mathfrak{F}_{t \phi, t} = 0.
\fe

In these equations, the vertical bar denotes intrinsic derivatives, while the comma represents directional derivatives with respect to the tetrad indices. By combining Eqs. \eqref{1e2d3e4m1} and \eqref{1e2d3e4m2} with the time derivative of Eq. \eqref{1e2d3e4m3}, the following result is therefore
\ie  \label{1e2d3e4m4}
\left[ \sqrt{A_{\Theta}(r) B^{-1}_{\Theta}(r)} \left( r \sqrt{A_{\Theta}(r)}\, \mathcal{F}
\right)_{,r} \right]_{,r} + \dfrac{A_{\Theta}(r) \sqrt{B_{\Theta}(r)}}{r} \left( \dfrac{%
\mathcal{F}_{,\theta}}{\sin\theta} \right)_{,\theta}\!\! \sin\theta - r 
\sqrt{B_{\Theta}(r)}\, \mathcal{F}_{,tt} = 0.
\fe
In this case, $\mathcal{F} = \mathfrak{F}_{t \phi } \sin\theta$. Notice that, by applying Fourier decomposition $(\partial_t \rightarrow - i \omega)$ and expressing the field in the form $\mathcal{F}(r,\theta) = \mathcal{F}(r) Y_{,\theta} / \sin\theta$, where $Y(\theta)$ corresponds to the Gegenbauer function \cite{g1,g2,g3,g5,g6}, Eq. \eqref{1e2d3e4m4} can be addressed 
\ie  \label{1e2d3e4m5}
\left[ \sqrt{A_{\Theta}(r) B^{-1}_{\Theta}(r)} \left( r \sqrt{A_{\Theta}(r)}\, \mathcal{F}
\right)_{,r} \right]_{,r} + \omega^2 r \sqrt{B_{\Theta}(r)}\, \mathcal{F} -
A_{\Theta}(r) \sqrt{B_{\Theta}(r)} r^{-1} \ell(\ell+1)\, \mathcal{F} = 0.
\fe

Introducing the redefinition \(\psi_\text{v} \equiv r \sqrt{A_{\Theta}(r)} \, \mathcal{F}\), Eq. \eqref{1e2d3e4m5} can be transformed into a Schrödinger--like equation. Under this formulation, it takes the form
\ie
\partial^2_{r_*} \psi_{\text{v}} + \omega^2 \psi_{\text{v}} = V^{(\text{v})}_{\text{eff}}(r,\Theta) \, \psi_{\text{v}},
\fe
which allows the effective potential for vector perturbations to be expressed as shown below
\ie  
\label{vectorpotential}
V^{(\text{v})}_{\text{eff}}(r,\Theta) = A_{\Theta}(r) \, \dfrac{\ell(\ell+1)}{r^2}.
\fe

\begin{figure}
    \centering
      \includegraphics[scale=0.519]{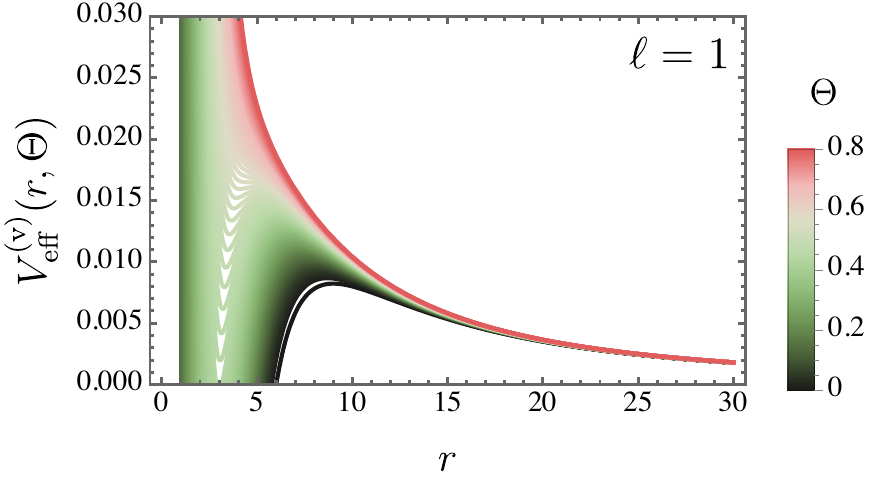}
      \includegraphics[scale=0.5]{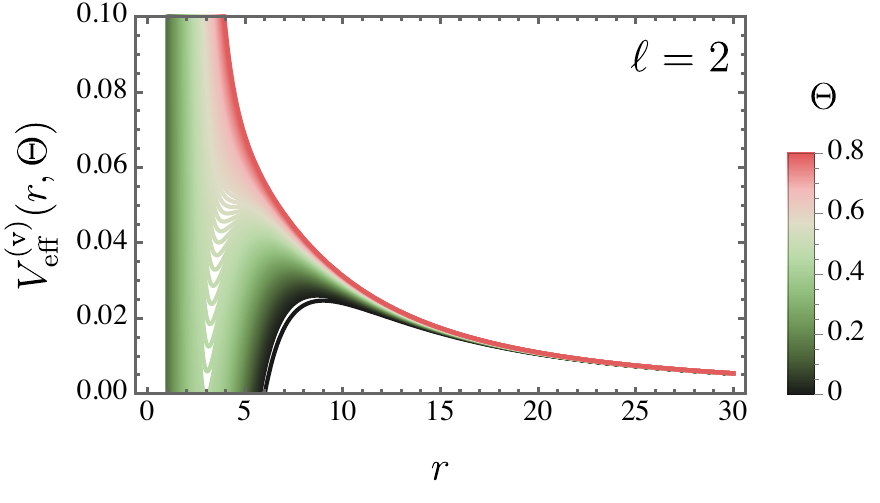}
      \includegraphics[scale=0.5]{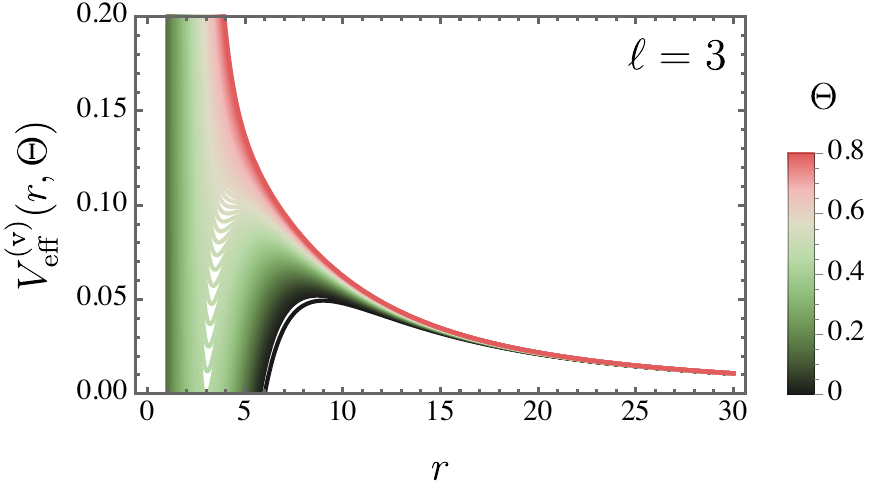}
    \caption{The vector effective potential, $V^{(\text{v})}_{\text{eff}}(r,\Theta)$, is presented for various values of the non--commutative parameter $\Theta$ and $\ell$.}
    \label{vs}
\end{figure}

By inserting Eq. (\ref{vectorpotential}) into Eq. (\ref{greybody}), the resulting expression is
\ie
\begin{split}
T_b^{\text{(v)}} = & \, \text{sech}^{2} \left(\int_{r_{ h}}^ {+\infty} \frac{V^{(\text{v})}_{\text{eff}}} {2\omega\sqrt{A_{\Theta}(r)B_{\Theta}(r)}}\mathrm{d}r\right) \\
= &  \, \text{sech}^{2} \left[ {\frac{1}{2\omega}\times} \frac{\sqrt[4]{\pi } \ell (\ell + 1) \sqrt{\sqrt{\pi } M-8 \sqrt{\Theta }}}{\sqrt{\pi } M^{3/2}+\sqrt[4]{\pi } M \sqrt{\sqrt{\pi } M-8 \sqrt{\Theta }}-8 \sqrt{\Theta  M}}  \right].
\end{split}
\fe

\begin{figure}
    \centering
      \includegraphics[scale=0.54]{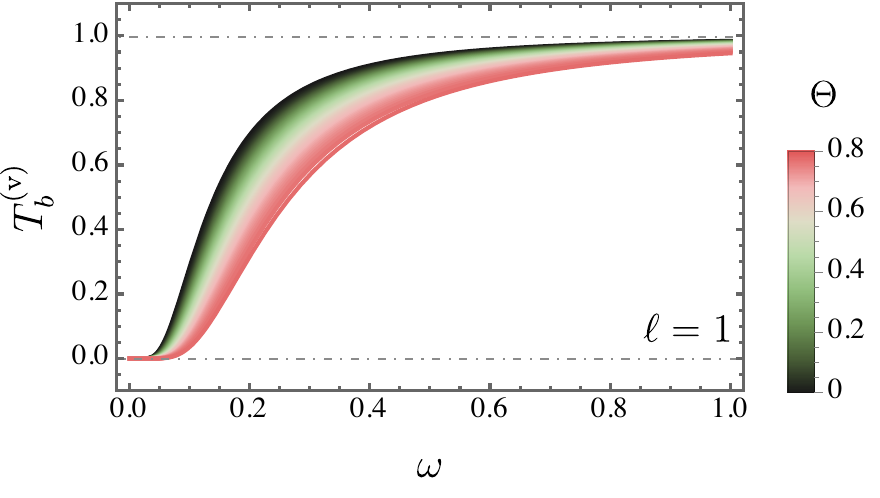}
      \includegraphics[scale=0.54]{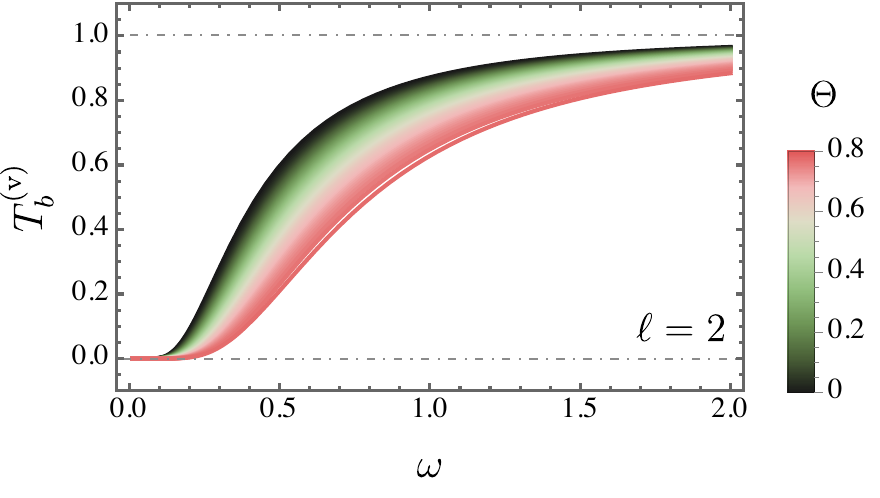}
      \includegraphics[scale=0.54]{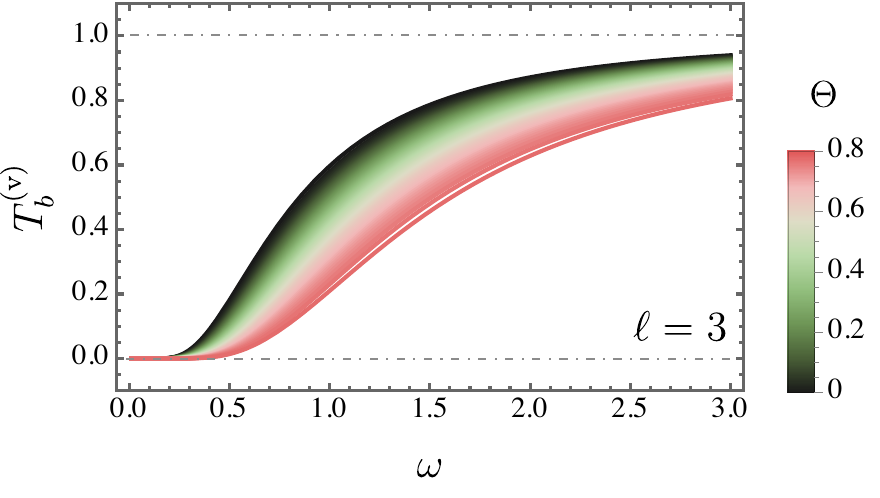}
    \caption{The greybody factors for vector perturbations, $T_{b}^{(\text{v})}$, are displayed for various values of the non--commutative parameter $\Theta$ and $\ell$.}
    \label{tbv}
\end{figure}

Fig. \ref{tbv} presents the greybody factors for vector perturbations, $T_{b}^{(\text{v})}$, for different values of the non--commutative parameter $\Theta$ and $\ell$. For the values of $\ell$ analyzed, an increase in $\Theta$ --- analogously to what happened to the scalar perturbations --- leads to a reduction in the magnitude of the greybody factors.

%%%%%%%%%%%%%%%%%%%%%%%%%%%%%%%%%%%%%%%%%%%%%%%%%%%%%%%%%%%%%%%%%%%%%%%%%%%%%%%%%%%%%%%%%%%%%%%%%%%%%%%%%%%%%%%%%%%%%%%%%%%%%%%%%%%%%%%%%%%%%%%%%%%%%%%%%%%%%%%%%%%%%%%%%%%%%%%%%%%%%%%%%%%%%%%%%%%%%%%%%%%%%%%%%%%%%%%%%%%%%%%%%%%%%%%%%%%%%%%%%%%%%%%%%%%%%%%%%%%%%%%%%%%%%%%%%%%%%%%%%%%%%%%%%%%%%%%%%%%%%%%%

\subsection{Tensorial perturbations}

This section focuses on investigating greybody factors in the context of tensorial (gravitational) perturbations. The analysis follows a methodology similar to that presented in Ref. \cite{kim2004gravitational,araujo2024dark,Filho:2024zxx}. Notably, a comparable approach can also be implemented using vierbein formalism \cite{Baruah:2025ifh,56Bouhmadi-Lopez:2020oia,59Chen:2019iuo}. To proceed, the axially symmetric spacetime is described by 
\ie
\mathrm{d}s^2 -e^{2\tilde{\nu}}\mathrm{d}t^2 + e^{2\tilde{\psi}}(\mathrm{d}\tilde{\phi} - \tilde{q}_1\mathrm{d}t - \tilde{q}_2\mathrm{d}r - \tilde{q}_3\mathrm{d}\theta )^2 e^{2\tilde{\mu}_2}\mathrm{d}r^2 + e^{2\tilde{\mu}_3}\mathrm{d}\theta^2. 
\fe
Considering the unperturbed black hole configuration, the following expression is obtained
\ie e^{2\tilde{\nu}} = A_{\Theta}(r), \quad
e^{-2\tilde{\mu}_2}=\left( 1 - \frac{2m(r)}{r} \right) = \frac{\tilde{\Delta}}{r^2}, 
\fe
with
\ie
\tilde{\Delta} = r^2 - 2 m(r) r, \quad e^{\tilde{\mu}_3} = r, \quad e^{\tilde{\psi}} = r
\sin\theta, 
\fe 
and 
\ie \tilde{q}_{1} = \tilde{q}_{2} = \tilde{q}_{3} = 0.
\fe
The metric given in Eq. \eqref{huashuas} can be rewritten in the following form:
\ie
A_{\Theta}(r)=1-\frac{2 \mathcal{M}_{\Theta}(r)}{r},
\fe
which leads to
\ie
\nonumber
\mathcal{M}_{\Theta}(r) = M - \frac{4M \sqrt{\Theta}}{\sqrt{\pi}r}.
\fe

The parameters $\tilde{q}_{1}$, $\tilde{q}_{2}$, and $\tilde{q}_{3}$ are commonly used to describe axial perturbations. Notably, when considering linear perturbations such as $\delta\tilde{\nu}$, $\delta\tilde{\psi}$, $\delta\tilde{\mu}_2$, and $\delta\tilde{\mu}_3$, even--parity polar modes arise. However, these modes will not be analyzed in this work. Shifting the focus to Einstein’s equations, the resulting expressions can be written as
\ie ( e^{3\tilde{\psi} + \tilde{\nu} - \tilde{\mu}_2 - \tilde{\mu}_3}
\tilde{Q}_{23} )_{,3} = - e^{3\tilde{\psi} - \tilde{\nu} -\tilde{\mu}_2 + \tilde{\mu}_3} \tilde{Q}_{02,0},\fe 
with
$x^2 = r, x^3 = \theta$ and $\tilde{Q}_{AB} = \tilde{q}_{A,B} - \tilde{q}_{B,A}, \tilde{Q}_{A0} =
\tilde{q}_{A,0} - \tilde{q}_{1,A}$ \cite{kim2004gravitational,araujo2024dark,Filho:2024zxx,Filho:2024zxx2}. Additionally, the above expression can be reformulated as
\ie
\frac{\sqrt{A_{\Theta}(r)}}{\sqrt{\tilde{\Delta}}}\frac{1}{r^3\sin^3\theta}\frac{\partial
\tilde{Q}}{\partial\theta} = - (\tilde{q}_{1,2} - \tilde{q}_{2,0})_{,0},
\fe 
in which $\tilde{Q}$ is given as
\ie
\tilde{Q}(t,r,\theta) = \tilde{\Delta} Q_{23}\sin^3\theta = \tilde{\Delta}
(\tilde{q}_{2,3} - \tilde{q}_{3,2})\sin^3\theta. \fe

Another key equation to consider is given by
\ie ( e^{3\tilde{\psi} + \tilde{\nu} - \tilde{\mu}_2 - \tilde{\mu}_3} \tilde{Q}_{23} )_{,2} =
e^{3\tilde{\psi} - \tilde{\nu} + \tilde{\mu}_2 - \tilde{\mu}_3} Q_{03,0}, \fe it can be shown that \ie
\frac{\sqrt{A_{\Theta}(r)}\sqrt{\tilde{\Delta}}}{r^3\sin^3\theta}\frac{\partial \tilde{Q}}{\partial\theta} =
(\tilde{q}_{1,3} - \tilde{q}_{3,0})_{,0}.\fe

This result can be further established by introducing the expression $ \tilde{Q}(r,\theta) = \tilde{Q}(r)C^{-3/2}_{l+2}(\theta)$, where $C^{\tilde{\nu}}_{n} (\theta)$ represents the Gegenbauer function, which satisfies the following relation \cite{kim2004gravitational,araujo2024dark,Filho:2024zxx}
\ie
\left[ \frac{\mathrm{d}}{\mathrm{d}\theta}\sin^{2\tilde{\nu}}\theta \frac{\mathrm{d}}{\mathrm{d}\theta} +
n(n + 2\tilde{\nu})\sin^{2\tilde{\nu}}\theta \right] C^{\tilde{\nu}}_n (\theta) = 0, \fe 
therefore
\ie r \sqrt{A_{\Theta}(r) \tilde{\Delta}} \frac{\mathrm{d}}{\mathrm{d}r} \left( \frac{\sqrt{A_{\Theta}(r) \tilde{\Delta}}
}{r^3} \frac{\mathrm{d}\tilde{Q}}{\mathrm{d}r} \right) - \tilde{\mu}^2
\frac{A_{\Theta}(r)}{r^2}\tilde{Q} + \omega^{2\tilde{Q}} = 0,  \fe 
in which
$\tilde{\mu}^2=(\ell-1)(\ell+2)$. Here, we define $\tilde{Q} = rZ$, which allows us to express the equation as
\ie
\label{eqqq}
\left( \frac{\mathrm{d}^2}{\mathrm{d}r^{*2}} +
\omega^2 -  V^{(\text{t})}_{\text{eff}}(r,\Theta) \right) Z = 0, \fe

Moreover, the effective potential can be derived as shown below
\ie
    V^{(\text{t})}_{\text{eff}}(r,\Theta) = A_{\Theta}(r) \left(\frac{\ell(\ell+1)}{r^2} - \frac{6 \mathcal{M}_{\Theta}(r)}{r^3}+\frac{2 \mathcal{M}_{\Theta}'(r)}{r^2} \right),
\fe
so that
\ie
\label{tensorpotential}
   V^{(\text{t})}_{\text{eff}}(r,\Theta) = A_{\Theta}(r) \left(\frac{\ell(\ell+1)}{r^2}  +  \frac{2 M \left(\frac{16 \sqrt{\Theta }}{\sqrt{\pi }}-3 r\right)}{r^4} \right).
\fe

\begin{figure}
    \centering
      \includegraphics[scale=0.519]{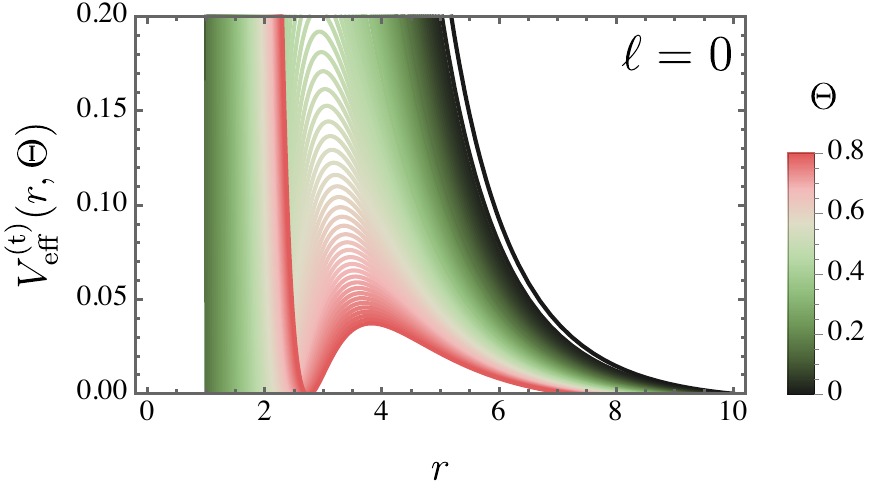}
      \includegraphics[scale=0.5]{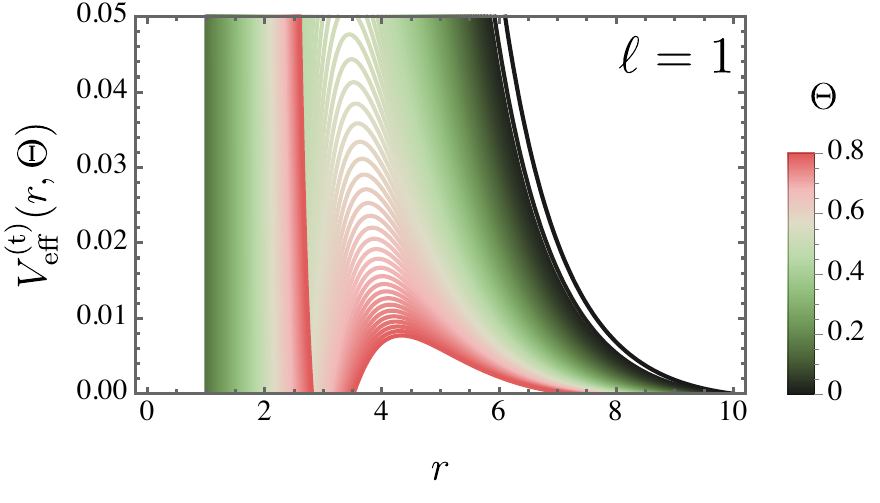}
      \includegraphics[scale=0.5]{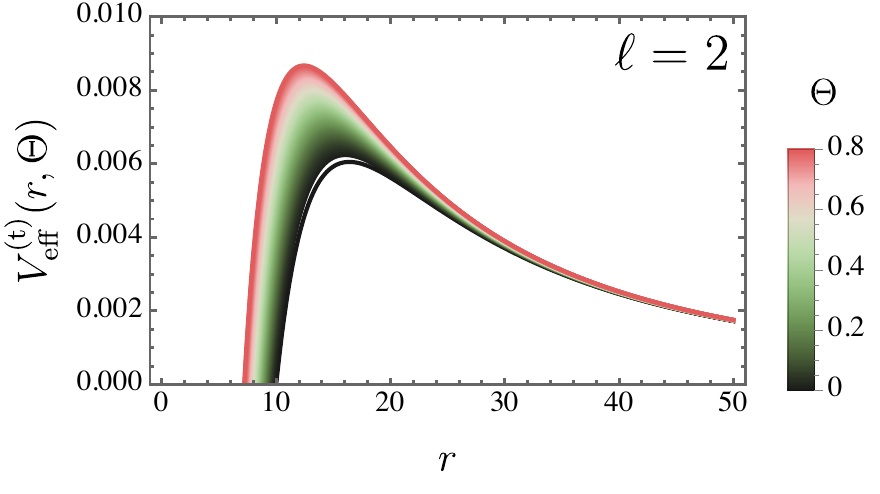}
    \caption{The tensorial effective potential, $V^{(\text{t})}_{\text{eff}}(r, \Theta)$, is presented for various values of the non--commutative parameter $\Theta$ and $\ell$.}
    \label{vt}
\end{figure}

Fig. \ref{vt} illustrates the behavior of $ V^{(\text{t})}_{\text{eff}}(r, \Theta)$ for various values of $\Theta$ and $\ell$. In general, when $\Theta$ increases, the effective potential decreases for $\ell = 0$. However, for $\ell = 1$ and $\ell = 2$, a larger $\Theta$ leads to an increase in the corresponding effective potential.

By inserting Eq. (\ref{tensorpotential}) into Eq. (\ref{greybody}), the resulting expression is given by
\ie
\begin{split}
T_b^{\text{(t)}} = & \, \text{sech}^{2} \left(\int_{r_{ h}}^ {+\infty} \frac{V^{(\text{t})}_{\text{eff}}} {2\omega\sqrt{A_{\Theta}(r)B_{\Theta}(r)}}\mathrm{d}r\right) \\
= &  \, \text{sech}^{2} \left\{ {\frac{1}{2\omega}\times} \frac{\sqrt[4]{\pi } M}{3 \left(\sqrt{M \left(\sqrt{\pi } M-8 \sqrt{\Theta }\right)}+\sqrt[4]{\pi } M\right)^3} \right. \\
& \left.  \times \left[ -8 \sqrt{\Theta } (3 \ell (\ell+1)-4)+3 \sqrt[4]{\pi } (2 l (\ell+1)-3) \sqrt{M \left(\sqrt{\pi } M-8 \sqrt{\Theta }\right)}  \right.\right. \\
& \left. \left. +3 \sqrt{\pi } (2 \ell (\ell+1)-3) M     \right]    \right\}.
\end{split}
\fe

\begin{figure}
    \centering
      \includegraphics[scale=0.54]{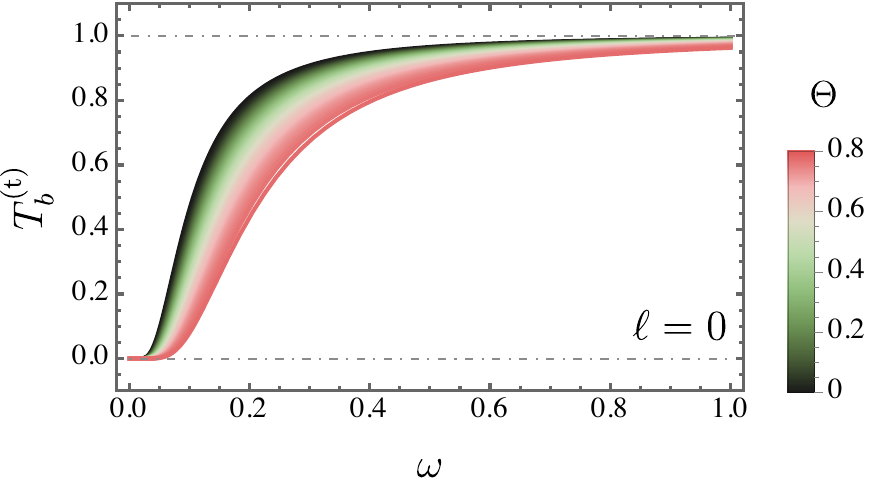}
      \includegraphics[scale=0.54]{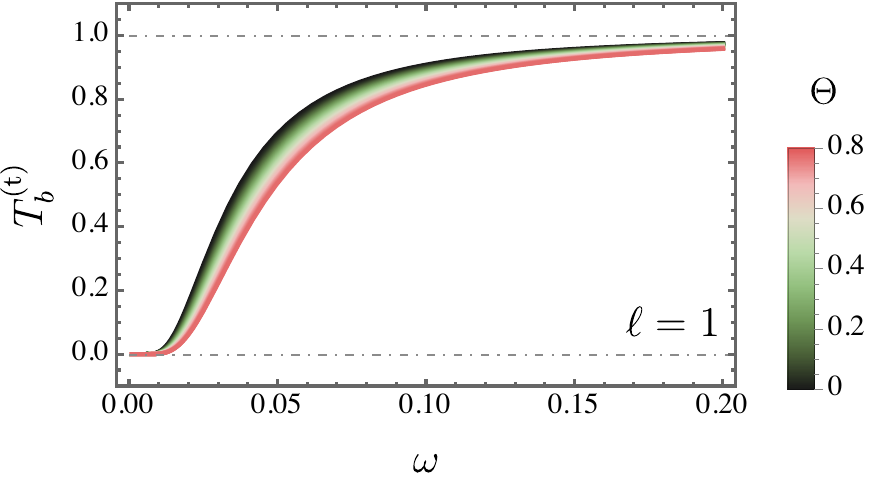}
      \includegraphics[scale=0.54]{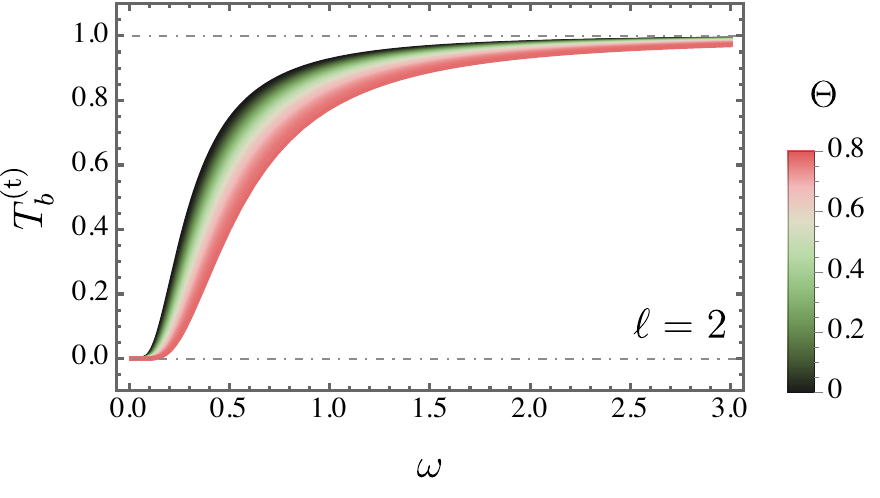}
    \caption{The greybody factors for tensor perturbations, $T_{b}^{(\text{t})}$, are displayed for various values of the non--commutative parameter $\Theta$ and $\ell$.}
    \label{tbt}
\end{figure}

Fig. \ref{tbt} presents the greybody factors for tensor perturbations, $T_{b}^{(\text{t})}$, for different values of the non--commutative parameter $\Theta$ and $\ell$. Analogouslly to the scalar and vector perturbations, an increase in $\Theta$ results in a reduction of the greybody factors' magnitude.

%%%%%%%%%%%%%%%%%%%%%%%%%%%%%%%%%%%%%%%%%%%%%%%%%%%%%%%%%%%%%%%%%%%%%%%%%%%%%%%%%%%%%%%%%%%%%%%%%%%%%%%%%%%%%%%%%%%%%%%%%%%%%%%%%%%%%%%%%%%%%%%%%%%%%%%%%%%%%%%%%%%%%%%%%%%%%%%%%%%%%%%%%%%%%%%%%%%%%%%%%%%%%%%%%%%%%%%%%%%%%%%%%%%%%%%%%%%%%%%%%%%%%%%%%%%%%%%%%%%%%%%%%%%%%%%%%%%%%%%%%%%%%%%%%%%%%%%%%%%%%%%%

\section{Greybody factors: fermionic case}

This section focuses on the analysis of massless Dirac perturbations in the spacetime of a static, spherically symmetric black hole. To investigate the behavior of the massless spin--$1/2$ field, we employ the Newman--Penrose formalism, which provides a convenient approach for describing fermionic dynamics in curved spacetime. The Dirac equations governing this system can be written as \cite{newman1962approach,chandrasekhar1984mathematical}
\begin{align}
(D + \epsilon - \rho) F_1 +( \bar{\delta} + \pi - \alpha) F_2 &= 0, \\
(\Delta + \mu - \gamma) F_2 + (\delta + \beta - \tau) F_1 &= 0,
\end{align}
in which $F_1$ and $F_2$ are the Dirac spinors, while the operators $D = l^\mu \partial_\mu$, $\Delta = n^\mu \partial_\mu$, $\delta = m^\mu \partial_\mu$, and $ \bar{\delta} = \bar{m}^\mu \partial_\mu$ denote the directional derivatives corresponding to the selected null tetrad frame.

For this analysis, the null tetrad basis vectors are formulated in terms of the metric components as
\begin{align}
l^\mu &= \left(\frac{1}{A}, \sqrt{\frac{B_{\Theta}(r)}{A_{\Theta}(r)}}, 0, 0\right), \quad \quad
n^\mu = \frac{1}{2} \left(1, -\sqrt{A_{\Theta}(r) B_{\Theta}(r)}, 0, 0\right), \\
m^\mu &= \frac{1}{\sqrt{2} r} \left(0, 0, 1, \frac{i}{\sin \theta}\right), \quad \quad
\bar{m}^\mu = \frac{1}{\sqrt{2} r} \left(0, 0, 1, \frac{-i}{\sin \theta}\right).
\end{align}

Based on these definitions, the non--zero components of the spin coefficients are determined below
\ie
 \rho = -\frac{1}{r} \frac{B_{\Theta}(r)}{A_{\Theta}(r)}, \quad \quad
\mu = -\frac{\sqrt{A_{\Theta}(r) B_{\Theta}(r)}}{2r},  \quad \quad
\gamma = \frac{A_{\Theta}(r)'}{4}\sqrt{\frac{B_{\Theta}(r)}{A_{\Theta}(r)}}, \quad \quad
\beta = -\alpha = \frac{\cot{\theta}}{2\sqrt{2}\, r} .
\fe

By decoupling the equations governing the massless Dirac field, a single equation of motion for \( F_1 \) is obtained
\begin{align}
\left[(D - 2\rho)(\Delta + \mu - \gamma) - (\delta + \beta) (\bar{\delta}+\beta)\right] F_1 = 0.
\end{align}

By incorporating the previously defined expressions for the directional derivatives and spin coefficients, the equation can be explicitly reformulated
\begin{align}
&\left[ \frac{1}{2A_{\Theta}(r)} \partial_t^2 - \left( \frac{\sqrt{A_{\Theta}(r)B_{\Theta}(r)}}{2r} +\frac{A_{\Theta}(r)'}{4}\sqrt{\frac{B_{\Theta}(r)}{A_{\Theta}(r)}}\right)\frac{1}{A_{\Theta}(r)}\partial_t \right. \\ \nonumber 
& \left. - \frac{\sqrt{A_{\Theta}(r)B_{\Theta}(r)}}{2} \sqrt{\frac{B_{\Theta}(r)}{A_{\Theta}(r)}}\partial_r^2 -\sqrt{\frac{B_{\Theta}(r)}{A_{\Theta}(r)}} \partial_r \left( \frac{\sqrt{A_{\Theta}(r)B_{\Theta}(r)}}{2} + \frac{A_{\Theta}(r)'}{4}{\sqrt{\frac{B_{\Theta}(r)}{A_{\Theta}(r)}}} \right) \right] F_1  \\ \nonumber
& + \left[ \frac{1}{\sin^2\theta} \partial_\phi^2 + i \frac{\cot \theta}{\sin \theta}\partial_\phi + \frac{1}{\sin \theta}\partial_\theta \left( \sin \theta \partial_\theta \right) - \frac{1}{4} \cot^2 \theta + \frac{1}{2} \right] F_1 = 0.
\end{align}

In order to decouple the equations into their radial and angular components, the wave function is represented as
\ie
F_1 = R(r) S_{l,m}(\theta, \phi) e^{-i \omega t},
\fe
with the radial component is expressed as
\begin{align}
&\left[\frac{-\omega^2}{2A_{\Theta}(r)} - \left(\frac{\sqrt{A_{\Theta}(r)B_{\Theta}(r)}}{2r}+\frac{A_{\Theta}(r)'}{4} + \sqrt{\frac{B_{\Theta}(r)}{A_{\Theta}(r)}}\right)\frac{- i\omega}{A_{\Theta}(r)} - \frac{\sqrt{A_{\Theta}(r)B_{\Theta}(r)}}{2} \sqrt{\frac{B_{\Theta}(r)}{A_{\Theta}(r)}}\partial_r^2 \right. \\ \nonumber
& \left. - \sqrt{\frac{B_{\Theta}(r)}{A_{\Theta}(r)}}\partial_r \left(\frac{\sqrt{A_{\Theta}(r)B_{\Theta}(r)}}{2r} + \frac{A_{\Theta}(r)'}{4}\sqrt{\frac{B_{\Theta}(r)}{A_{\Theta}(r)}}\right)-\lambda_l\right] R(r) = 0.
\end{align}

In this context, $\lambda_{l}$ functions as the separation constant. Recasting the radial coordinate in terms of the generalized tortoise coordinate $r^{*}$, as specified in Eq. \eqref{rstar}, allows the radial equation to take the form of a Schrödinger--like wave equation, given by:
\begin{align}
\left[\frac{\mathrm{d}^2 }{\mathrm{d}r_*^2} +( \omega^2 - V_{\text{eff}}^{\uparrow \downarrow}) \right]U_{\uparrow \downarrow} = 0.
\end{align}
The effective potentials $V_{\text{eff}}^{\uparrow \downarrow}$ corresponding to the massless spin--$1/2$ field are expressed as follows \cite{albuquerque2023massless, al2024massless, arbey2021hawking}
\ie
\label{Vpm}
V_{\text{eff}}^{\uparrow \downarrow} = \frac{(\ell + \frac{1}{2})^2}{r^2} A_{\Theta}(r) \pm \left(\ell + \frac{1}{2}\right) \sqrt{A_{\Theta}(r) B_{\Theta}(r)} \partial_r \left(\frac{\sqrt{A_{\Theta}(r)}}{r}\right).
\fe

Without any loss of generality, we choose $V_{\text{eff}}^{\uparrow}$ as the potential for analysis. A similar treatment applies to $V_{\text{eff}}^{\downarrow}$; however, since its behavior closely resembles that of $V_{\text{eff}}^{\uparrow}$ \cite{albuquerque2023massless,devi2020quasinormal}, our discussion will be centered on $V_{\text{eff}}^{\uparrow}$. In an explicitly manner, we have:
\ie
V^{\uparrow}_{\text{eff}}(r,\Theta) = \frac{(2 {\ell} +1) \left(\sqrt{\pi } r (3 M-r)-16 \sqrt{\Theta } M\right)}{2 \pi ^{3/4} r^5 \sqrt{\frac{1}{8 \sqrt{\Theta } M+\sqrt{\pi } r (r-2 M)}}}+\frac{\left({\ell}+\frac{1}{2}\right)^2 \left(\frac{2 M \left(\frac{4 \sqrt{\Theta }}{\sqrt{\pi }}-r\right)}{r^2}+1\right)}{r^2}.
\fe

\begin{figure}
    \centering
      \includegraphics[scale=0.54]{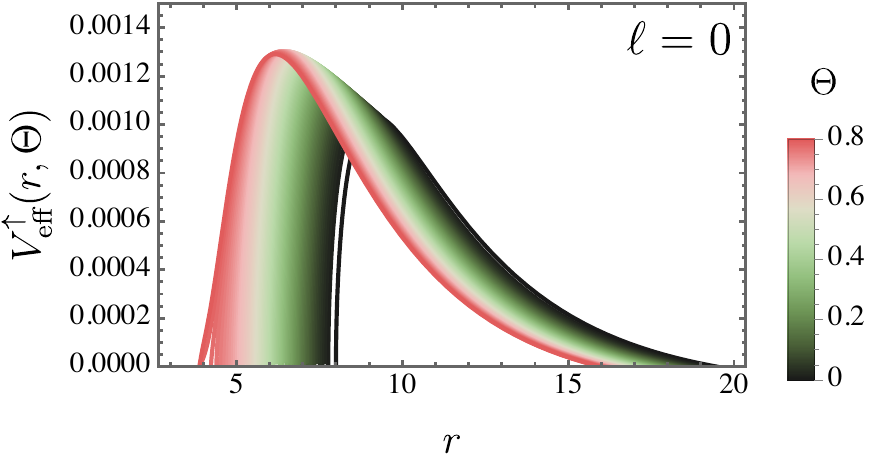}
      \includegraphics[scale=0.54]{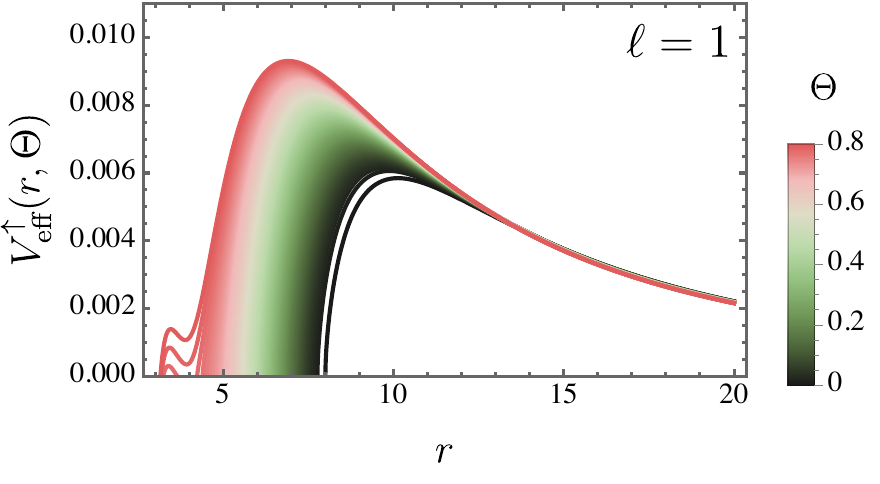}
      \includegraphics[scale=0.54]{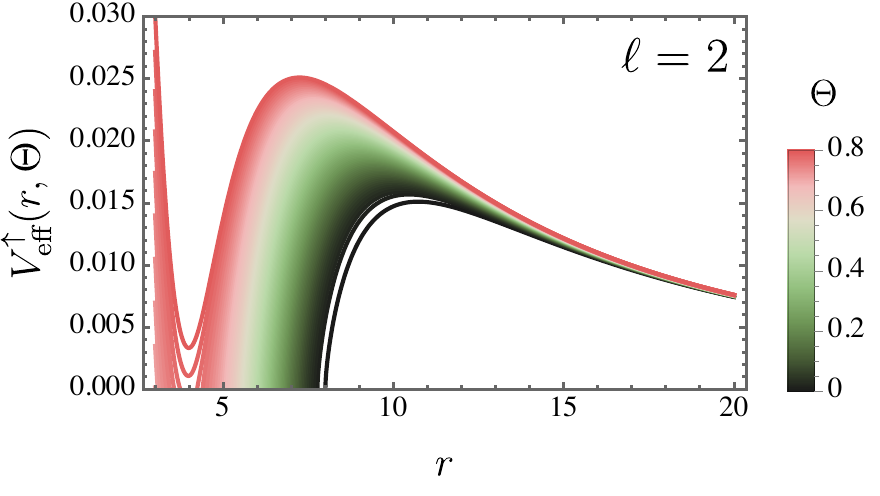}
    \caption{The effective potential for the fermionic case, $V_{\text{eff}}^{\uparrow}$, is shown for different values of $\Theta$ and $\ell$.}
    \label{vfermions}
\end{figure}

To better interpret the above expression, Fig. \ref{vfermions} illustrates the behavior of $V^{\uparrow}_{\text{eff}}(r,\Theta)$ for different values of $\Theta$ and $\ell$. For $\ell = 1$ and $\ell = 2$, the overall trend shows that as $\Theta$ increases, the potential also increases.  As a result, the greybody factors are given by:
\ie
\begin{split}
T_b^{\uparrow} = & \, \text{sech}^{2} \left(\int_{r_{ h}}^ {+\infty} \frac{V^{\uparrow}_{\text{eff}}} {2\omega\sqrt{A_{\Theta}(r)B_{\Theta}(r)}}\mathrm{d}r\right).
\end{split}
\fe

\begin{figure}
    \centering
      \includegraphics[scale=0.51]{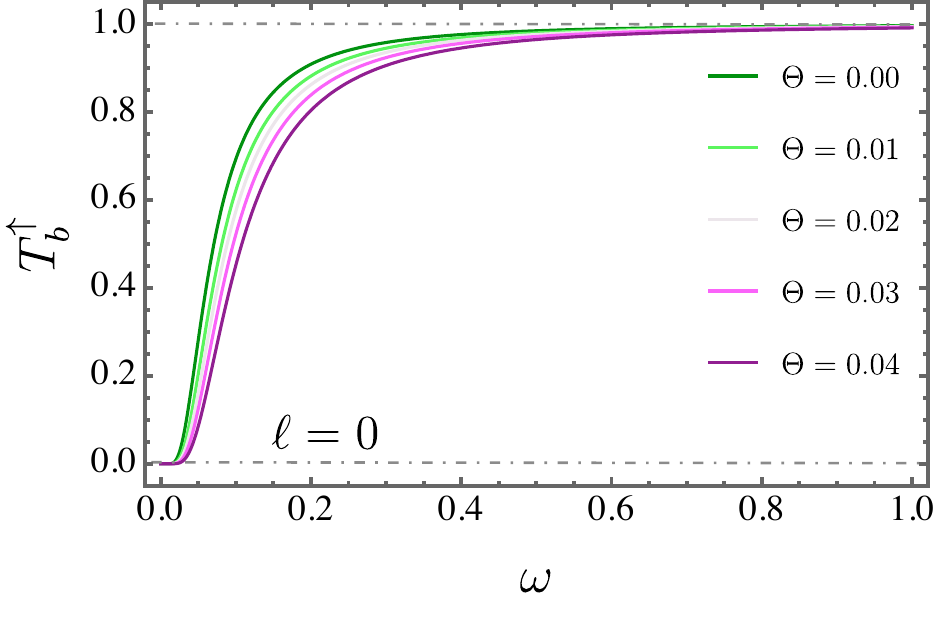}
      \includegraphics[scale=0.51]{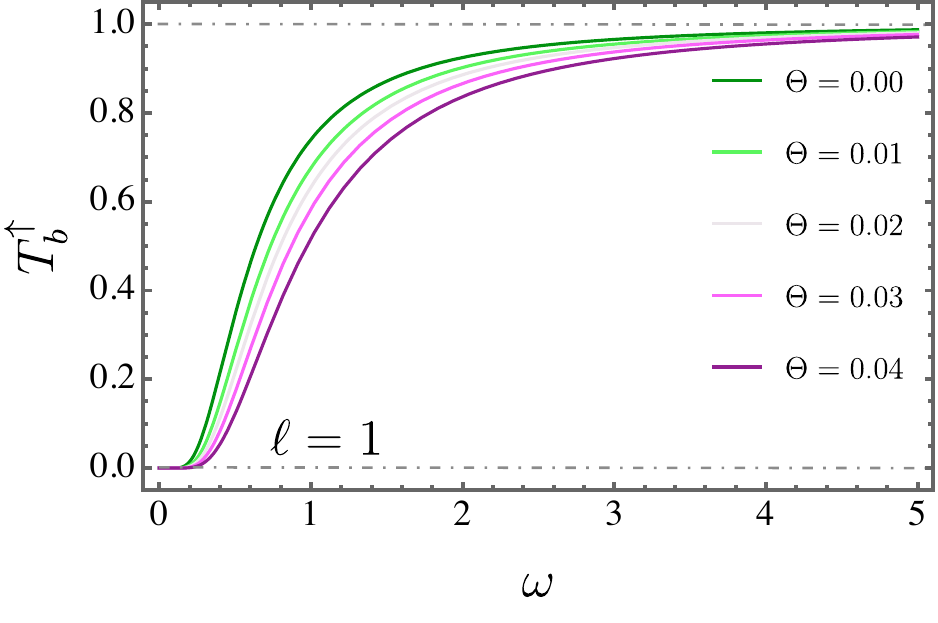}
      \includegraphics[scale=0.51]{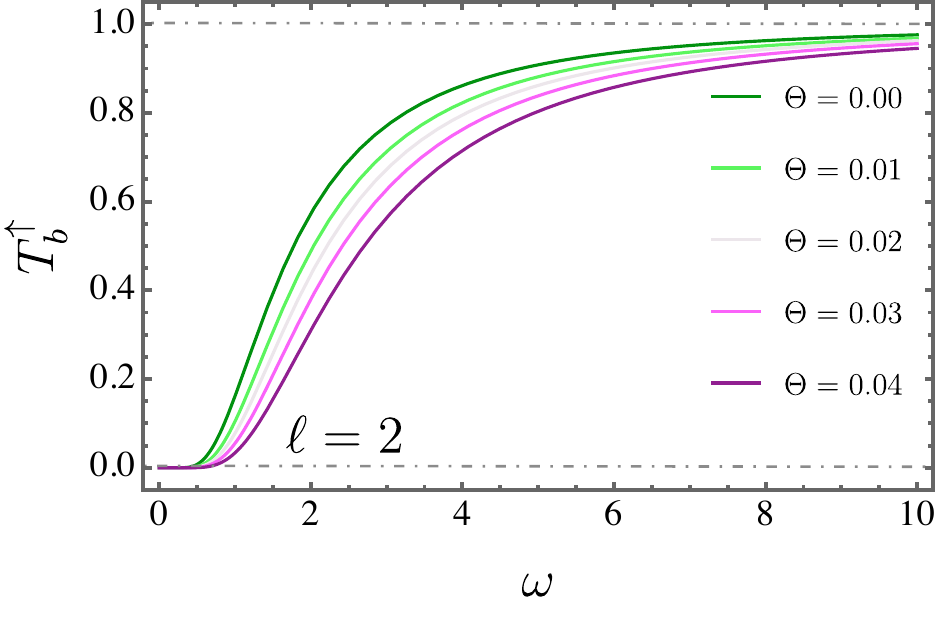}
    \caption{The effective potential for the fermionic case $V_{\text{eff}}^{\uparrow}$ is shown for different values of $\Theta$ and $\ell$.}
    \label{tbfermions}
\end{figure}

It is important to note that, unlike the greybody factor calculations in previous sections, the present analysis is carried out numerically. The results are displayed in Fig. \ref{tbfermions}, where they are compared to the Schwarzschild case ($\Theta = 0$). In general, the parameter $\Theta$ leads to a reduction in the magnitude of $T_b^{\uparrow}$. Compared to the Schwarzschild case, which exhibits the highest magnitude, the results show a noticeable attenuation.

%%%%%%%%%%%%%%%%%%%%%%%%%%%%%%%%%%%%%%%%%%%%%%%%%%%%%%%%%%%%%%%%%%%%%%%%%%%%%%%%%%%%%%%%%%%%%%%%%%%%%%%%%%%%%%%%%%%%%%%%%%%%%%%%%%%%%%%%%%%%%%%%%%%%%%%%%%%%%%%%%%%%%%%%%%%%%%%%%%%%%%%%%%%%%%%%%%%%%%%%%%%%%%%%%%%%%%%%%%%%%%%%%%%%%%%%%%%%%%%%%%%%%%%%%%%%%%%%%%%%%%%%%%%%%%%%%%%%%%%%%%%%%%%%%%%%%%%%%%%%%%%%%%%%%%%%%%%%%%%%%%%%%%%%%%%%%%%%%%%%%%%%%%%%%%%%%%%%%%%%%%%%%%%%%%%%%%%%%%%%%%%%%%%%%%%%%%%%%%%%%%%%%%%%%%%%%%%%%%%%

\section{Conclusion}

In this paper, we investigated particle production, evaporation, and greybody factors for a Lorentzian non--commutative black hole. We began by analyzing bosonic particle creation, considering scalar perturbations to compute the Bogoliubov coefficients, $\alpha^{\Theta}_{\omega\omega^\prime}$ and $\beta^{\Theta}_{\omega\omega^\prime}$, which allowed us to determine the Hawking temperature, $T_{\Theta}$. The non--commutative parameter $\Theta$ was found to suppress its magnitude.  

Next, we described Hawking radiation as a tunneling process using the Painlevé--Gullstrand metric representation, so that we could evaluate the divergent integrals via the residue method. This approach led to the derivation of the particle creation density for bosonic modes, $n_{b}(\omega,\Theta)$. We then extended the analysis to fermions, obtaining the corresponding density $n_{f}(\omega,\Theta)$. In both cases, the presence of $\Theta$ consistently reduced the particle production rates.  

The black hole evaporation was examined within the framework of the \textit{Stefan--Boltzmann} law, leading to an estimate of its lifetime in the high--frequency limit, where radiation was well approximated by null geodesics in the geometrical optics regime. As $\Theta$ increased, the evaporation process slowed down, as reflected in the final evaporation time, $t_{evapfinal}$. Additionally, we identified the emergence of a remnant mass, $M_{\text{rem}}$, when the black hole reached the final stage of its evaporation.  

Furthermore, we computed greybody factors for bosonic fields, considering scalar ($T_b^{\text{(s)}}$), vector ($T_b^{\text{(v)}}$), and tensorial ($T_b^{\text{(t)}}$) perturbations. We also determined the greybody factors for fermionic fields, denoted by $T_b^{\uparrow}$.  For all these configurations, $\Theta$ turned out to reduce the magnitude of them.

As a natural extension, it would be worthwhile to perform a similar analysis for different implementations of non--commutativity, such as a Gaussian distribution, and compare the results. This and other related investigations are currently in progress.

%%%%%%%%%%%%%%%%%%%%%%%%%%%%%%%%%%%%%%%%%%%%%%%%%%%%%%%%%%%%%%%%%%%%%%%%%%%%%%%%%%%%%%%%%%%%%%%%%%%%%%%%%%%%%%%%%%%%%%%%%%%%%%%%%%%%%%%%%%%%%%%%%%%%%%%%%%%%%%%%%%%%%%%%%%%%%%%%%%%%%%%%%%%%%%%%%%%%%%%%%%%%%%%%%%%%%%%%%%%%%%%%%%%%%%%%%%%%%%%%%%%%%%%%%%%%%%%%%%%%%%%%%%%%%%%%%%%%%%%%%%%%%%%%%%%%%%%%%%%%%%%%%%%%%%%%%%%%%%%%%%%%%%%%%%%%%%%%%%%%%%%%%%%%%%%%%%%%%%%%%%%%%%%%%%%%%%%%%%%%%%%%%%%%%%%%%%%%%%%%%%%%%%%%%%%%%%%%%%%%

\section*{Acknowledgments}
\hspace{0.5cm} A. A. Araújo Filho is supported by Conselho Nacional de Desenvolvimento Cient\'{\i}fico e Tecnol\'{o}gico (CNPq) and Fundação de Apoio à Pesquisa do Estado da Paraíba (FAPESQ), project No. 150891/2023-7.

\bibliographystyle{ieeetr}
\bibliography{main}

\end{document}